\documentclass[a4paper,11pt]{article}
\pdfoutput=1
\usepackage{jheppub}

\usepackage{amssymb,amsmath}
\usepackage[normalem]{ulem}
\usepackage[utf8x]{inputenc}
\usepackage{slashed}
\usepackage{graphicx}
\usepackage{here}
\usepackage{color}
\usepackage{csquotes} 
\usepackage{comment}
\usepackage{mathrsfs}
\usepackage{float}
\usepackage{ascmac}
\usepackage{multirow}
\usepackage{longtable}
\usepackage{bm}

\usepackage[italicdiff]{physics}

\usepackage[hang,small,bf]{caption}
\usepackage[subrefformat=parens]{subcaption}
\makeatletter
\newcommand*\rel@kern[1]{\kern#1\dimexpr\macc@kerna}
\newcommand*\widebar[1]{%
  \begingroup
  \def\mathaccent##1##2{%
    \rel@kern{0.8}%
    \overline{\rel@kern{-0.8}\macc@nucleus\rel@kern{0.2}}%
    \rel@kern{-0.2}%
  }%
  \macc@depth\@ne
  \let\math@bgroup\@empty \let\math@egroup\macc@set@skewchar
  \mathsurround\z@ \frozen@everymath{\mathgroup\macc@group\relax}%
  \macc@set@skewchar\relax
  \let\mathaccentV\macc@nested@a
  \macc@nested@a\relax111{#1}%
  \endgroup
}
\makeatother

\numberwithin{equation}{section}

\preprint{
\begin{minipage}{5cm}
\small
\flushright
EPHOU-24-010\\
KYUSHU-HET-292
\end{minipage}}

\title{Non-invertible flavor symmetries in magnetized extra dimensions}

\author{Tatsuo Kobayashi$^{1}$ and} 
\author{Hajime Otsuka$^{2}$} 
\affiliation{
$^1$Department of Physics, Hokkaido University, Sapporo 060-0810, Japan}
\affiliation{
$^2$Department of Physics, Kyushu University, 744 Motooka, Nishi-ku, Fukuoka 819-0395, Japan}
\emailAdd{kobayashi@particle.sci.hokudai.ac.jp}
\emailAdd{otsuka.hajime@phys.kyushu-u.ac.jp}

\abstract{
We discuss non-invertible symmetries in toroidal compactifications of higher dimensional Yang-Mills theory with magnetic fluxes, which can be regarded as an effective action of type IIB string theory with magnetized D-branes. 
Specifically, we derive fusion rules of discrete isometry operators, which are invariant under the $\mathbb{Z}_N$ orbifold twist. 
It turns out that degenerate chiral zero-modes non-trivially transform under the non-invertible symmetries in the context of quantum mechanics. 
Hence, non-invertible symmetries correspond to flavor symmetries of chiral zero modes and determine their flavor structure. 
We explicitly show the representations and selection rules of chiral zero modes on their wave functions. These zero modes may correspond to generations of quarks and leptons in four-dimensional low-energy effective field theory. 
}

\makeatletter
\gdef\@fpheader{}
\makeatother

\begin{document}

\maketitle

\section{Introduction}

Symmetry is quite important in physics.
Recently, the concept of symmetries was generalized in the sense that 
an existence of topological operators corresponds to the symmetry in this system \cite{Gaiotto:2014kfa}. 
Furthermore, they provide us with new sights of symmetries in particle physics and others.

Many examples with non-invertible symmetries have been studied in field theory such as Verlinde lines in two-dimensional conformal field theories \cite{Verlinde:1988sn,Moore:1988qv,Moore:1989yh} (see, e.g., Refs.~\cite{Schafer-Nameki:2023jdn,Shao:2023gho} about reviews on non-invertible symmetries in various dimensions). 
For example, a non-invertible symmetry controls $n$-point coupling of twist fields on $S^1/\mathbb{Z}_2$ \cite{Heckman:2024obe,Kaidi:2024wio}. 
From their fusion rules, the $D_4$ symmetry was shown in Ref.~\cite{Kaidi:2024wio}. 
Such $D_4$ symmetry in the conformal field theory on $S^1/\mathbb{Z}_2$ 
was derived in Ref.~\cite{Dijkgraaf:1987vp}, and 
it was rediscovered as a flavor symmetry of matter fields 
in four-dimensional (4D) low-energy effective field theory 
of heterotic string theory on orbifold compactifications \cite{Kobayashi:2004ya,Kobayashi:2006wq}.
These flavor symmetries are important to understand 
the origin of fermion mass hierarchies and their mixing angles \cite{Altarelli:2010gt,Ishimori:2010au,Kobayashi:2022moq,Hernandez:2012ra,King:2013eh}.

The non-invertible symmetry is a peculiar 
symmetry in orbifold compactifications. 
As discussed in Ref.~\cite{Heckman:2024obe}, a topological operator including a translation in two-dimensional conformal field theory is invertible on the $T^2$ torus, but it is not a gauge invariant quantity under an orbifold group $\Gamma$ of isometries of the underlying torus. It restricts the form of the operator on toroidal orbifolds. 
In this respect, the existence of non-invertible symmetry is quite natural in orbifold compactifications. 
In this paper, we focus on type IIB string theory on toroidal orbifold compactifications with magnetized D-branes. 
Since magnetized extra dimensions provide us with similar matter flavor symmetries 
\cite{Abe:2009vi,Berasaluce-Gonzalez:2012abm,Marchesano:2013ega}, 
one can expect that a non-invertible symmetry appears in this system similarly to heterotic orbifold models.

Our purpose is to give new examples of field theory where non-invertible symmetries appear and study its meaning as flavor symmetries of chiral matters. 
In magnetized extra dimensions, background magnetic fluxes induce degenerate chiral zero modes whose degeneracy is determined by an amount of background fluxes \cite{Witten:1984dg}. 
So far, flavor symmetries discussed in magnetized extra dimensions are invertible for a given flavor group. 
However, in toroidal orbifold compactifications such as $T^2/\mathbb{Z}_N$, one can construct a momentum operator as well as matter wave functions in a $\mathbb{Z}_N$-invariant way. 
When degenerate chiral zero-modes non-trivially transform under the non-invertible symmetries, non-invertible symmetries will determine the flavor structure of chiral zero modes. 
In this paper, we explicitly show the fusion rules of $\mathbb{Z}_N$-invariant momentum operators on $T^2/\mathbb{Z}_N$ with magnetic fluxes. 
Furthermore, representations of chiral zero modes under the non-invertible symmetry are derived for a given magnetic flux. 
We do not explore realistic string compactifications, but when these chiral zero modes are identified with quarks and leptons, flavor symmetries originating from non-invertible symmetries will determine the flavor structure of the Standard Model.

This paper is organized as follows. 
In Sec. \ref{sec:2}, we review the magnetized D-brane models in type IIB toroidal compactifications with an emphasis on translations and flavor symmetries. 
In the end of Sec. \ref{sec:2}, we introduce a quantum mechanical description of the magnetized D-brane system. 
In Sec. \ref{sec:3}, we introduce the non-invertible symmetry in $T^2/\mathbb{Z}_2$ orbifold in the context of quantum mechanics. 
Then, the fusion rule and non-invertible transformations of chiral zero modes are explicitly shown. 
Such an analysis is extended to $T^2/\mathbb{Z}_N$ orbifolds with $N=3,4,6$  in Sec. \ref{sec:4}. 
Sec. \ref{sec:con} is devoted to conclusions. 
In Appendix \ref{app}, we summarize the transformation of chiral zero modes under the non-invertible symmetry on $T^2/\mathbb{Z}_N$ with $N=3,4,6$.

\section{Non-Abelian flavor symmetries on $T^2$}
\label{sec:2}

In this section, we review the non-Abelian flavor symmetries on $T^2$ in the framework of type IIB string theory on $T^6$ with magnetized D-branes. 
After discussing the translations in magnetized extra dimensions in Sec. \ref{sec:translationsT2}, we present zero-mode wave functions and 
selection rules in Sec. \ref{sec:flavorT2}. 
In Sec. \ref{sec:operator}, we rewrite the magnetized D-brane system in the context of quantum mechanics, which is useful 
for analyzing non-invertible symmetries on the orbifold. 

\subsection{Translations and non-Abelian groups}
\label{sec:translationsT2}

We start with type IIB string theory on $T^6= (T^2)_1\times (T^2)_2\times (T^2)_3$. 
Each two-dimensional torus is defined as $(T^2)_i= \mathbb{C}/\Lambda_i$ with a two-dimensional 
lattice $\Lambda_i$ ($i=1,2,3)$. 
When we choose the coordinate of each torus $(T^2)_i$ as $(y_{2i-1}, y_{2i})$ with $\{y_{2i-1}, y_{2i}\} \in \mathbb{R}$, 
there are six isometries: 
\begin{align}
    y_{2i-1} \rightarrow y_{2i-1} + \lambda_{y_{2i-1}},\qquad
    y_{2i} \rightarrow y_{2i} + \lambda_{y_{2i}}.    
    \label{eq:T6iso}
\end{align}
In the 4D effective theory, they correspond to $U(1)^6$ gauge symmetry whose gauge bosons are 
denoted as $(V_\mu^{y_{2i-1}}, V_\mu^{y_{2i}})$ in what follows. 
Furthermore, there are additional $U(1)^6$ gauge symmetries arising from the toroidal compactification 
of B-field whose gauge bosons are denoted as $(B_\mu^{y_{2i-1}}, B_\mu^{y_{2i}})$.

To make our analysis concrete, we consider magnetized D-branes, whose low-energy effective theory is described by 
$U(N)$ supersymmetric Yang-Mills action. 
In particular, we first focus on two stacks of magnetized D-branes $a$ and $b$, wrapping a $(T^2)_1$. 
They have $U(2)$ gauge symmetry if magnetic flux background vanishes. 
A complex coordinate of $(T^2)_1$ is represented as $z = y_1+ \tau y_2$ with $\tau$ being the complex structure. 
By introducing $U(1)$ magnetic flux along the coordinate of $T^2$:
\begin{align}
    F = \frac{\pi i}{{\rm Im}\,(\tau)} (m_a X_a + m_b X_b)
    dz \wedge d\Bar{z},
\end{align}
with 
\begin{align}
        X_a =
        \begin{pmatrix}
        1 & 0\\
        0 & 0 \\
    \end{pmatrix}
    ,\qquad
        X_b =
        \begin{pmatrix}
        0 & 0\\
        0 & 1 \\
    \end{pmatrix}
    ,
\end{align}
the original $U(2)$ gauge symmetry is broken down to $U(1)_a\times U(1)_b$ if the quantized fluxes are different from each other, i.e., $m_a \neq m_b$. 
The corresponding $U(1)$ vector potential is described by
\begin{align}
    A(y_1, y_2) = \pi (m_a X_a + m_b X_b)(y_1 dy_2 - y_2 dy_1).
\end{align}
Under Eq.~\eqref{eq:T6iso}, the vector potential transforms as
\begin{align}
    A(y_1 +\lambda_{y_1}, y_2) = A(y_1, y_2) + d\chi_1\,,\qquad
    A(y_1, y_2 +\lambda_{y_2}) = A(y_1, y_2) + d\chi_2\,,\qquad
\label{eq:Agauge}
\end{align}
with
\begin{align}
    \chi_1 &= \pi \lambda_{y_1} (m_a X_a + m_b X_b) dy_2\,,
    \qquad
    \chi_2 = - \pi \lambda_{y_2} (m_a X_a + m_b X_b) dy_1\,.
\end{align}
In this context, the translations are generated by $e^{\lambda_{y_1}D_{y_1}}$ and $e^{\lambda_{y_2}D_{y_2}}$ with $D_{y_1,y_2}$ being the covariant derivative, which satisfy the following relation:
\begin{align}
    e^{\lambda_i D_i}(iD_j)e^{-\lambda_i D_i} = i D_j + \lambda_i F_{ij}\,,
\label{eq:lambda_alg}
\end{align}
with $i,j=y_1,y_2$ and $[D_{y_1}, D_{y_2}] = -i F_{y_1y_2}$. 
In addition, when we turn on the B-field along 
the coordinate of $T^2$: 
\begin{align}
    B_x = 2\pi i y_1\,,\qquad
    B_y = 2\pi i y_2\,,
\end{align}
the gauge covariant derivatives satisfy 
\begin{align}
    e^{\mu_i D_i}(iD_j)e^{-\mu_i D_i} = i D_j + \mu_i \delta_{ij}\,.
\end{align}
By combining them, we arrive at 
\begin{align}
    e^{\lambda_{y_1} D_{y_1} + \mu_{y_2} B_{y_2}}(iD_i)e^{-\lambda_{y_1} D_{y_1} -\mu_{y_2} B_{y_2}} &= e^{2\pi i (\xi_{y_2,a}X_a + \xi_{y_2,b}X_b)y_2}(i D_i)e^{-2\pi i (\xi_{y_2,a}X_a + \xi_{y_2,b}X_b)y_2}\,,
    \nonumber\\
    e^{\lambda_{y_2} D_{y_2} + \mu_{y_1} B_{y_1}}(iD_i)e^{-\lambda_{y_2} D_{y_2} -\mu_{y_1} B_{y_1}} &= e^{2\pi i (\xi_{y_1,a}X_a + \xi_{y_1,b}X_b)y_1}(i D_i)e^{-2\pi i (\xi_{y_1,a}X_a + \xi_{y_1,b}X_b)y_1}\,, 
\end{align}
with $i=1,2$ and
\begin{align}
    \xi_{y_1,a} &= -\lambda_{y_2} m_a + n_a \mu_{y_1}\,,\qquad
    \xi_{y_1,b} = -\lambda_{y_2} m_b + n_b \mu_{y_1}\,,
    \nonumber\\
    \xi_{y_2,a} &= \lambda_{y_1} m_a + n_a \mu_{y_2}\,,\qquad
    \xi_{y_2,b} = \lambda_{y_1} m_b + n_b \mu_{y_2}\,,
\end{align}
which correspond to the Wilson-line scalars of $U(1)_a$ and $U(1)_b$.
Since the background gauge field is not invariant under an arbitrary translation, $U(1)^2$ isometries are in general broken down. However, as pointed out in Ref.~\cite{Marchesano:2013ega}, a certain choice of $\lambda_{x,y}$ allows us to keep a discrete symmetry in 4D effective theory. 
Indeed, if $\{\xi_{y_1,a}, \xi_{y_1,b}\} \in \mathbb{Z}^2$ and $\{\xi_{y_2,a}, \xi_{y_2,b}\} \in \mathbb{Z}^2$, the gauge transformations are manifest. 
There are $M=m_a- m_b$ choices for $\{\lambda_{y_1}, \mu_{y_2}\} \in \mathbb{Z}^2$ and $\{\lambda_{y_2}, \mu_{y_1}\} \in \mathbb{Z}^2$, corresponding to two $\mathbb{Z}_M$ symmetries. However, they do not commute with each other due to Eq.~\eqref{eq:lambda_alg}, which results in a non-Abelian group, namely the so-called Heisenberg group $H_M\simeq (\mathbb{Z}_M \times \mathbb{Z}_M) \rtimes \mathbb{Z}_M$.

\subsection{Zero-mode wave functions and selection rules}
\label{sec:flavorT2}

The discrete symmetry arising from the translation acts on the chiral zero modes which belong to the bifundamental representation of $U(1)_a\times U(1)_b$ with the charge (1,-1). 
It was known that the mode function of fermion zero-modes is described by~\cite{Cremades:2004wa}
\begin{align}\label{eq:wf}
    \psi^{j,M}(\tau, z) = {\cal N}_j e^{i\pi M \frac{z {\rm Im}\,(z)}{{\rm Im}\,(\tau)}}
    \vartheta\,
\begin{bmatrix}
\frac{j}{M}\\
0
\end{bmatrix}
(Mz, M\tau)
\end{align}
with the normalization factor
\begin{align}
    {\cal N}_j = \left(\frac{2 {\rm Im}\,(\tau)M}{{\cal A}^2}\right)^{1/4}\,,
\end{align}
for all $j = 0,1,...,M-1$, where $\cal A$ denotes the area of $T^2$. 
Here and in what follows, we focus on the $M>0$ case, and $\vartheta$ denotes the Jacobi theta function as 
a function of the complex structure modulus $\tau$:
\begin{align}
\vartheta
\begin{bmatrix}
a\\
b
\end{bmatrix}
(cz, c\tau)
\equiv 
\sum_{l\in \mathbb{Z}}
e^{\pi i (a+l)^2c\tau}e^{2\pi i (a+l)(cz +b)}
\,.
\end{align}
Note that the charged fermion, i.e., the bifundamental representation of $U(1)_a \times U(1)_b$ with charge $(1,-1)$, obeys the pseudo-periodic boundary condition:
\begin{align}
    \psi(y_1,y_2) &\rightarrow e^{-i \chi_1} \psi (y_1 + \lambda_{y_1}, y_2) = e^{\lambda_{y_1} D_{y_1}}\psi(y_1, y_2)\,,
    \nonumber\\
    \psi(y_1,y_2) &\rightarrow e^{-i \chi_2} \psi (y_1, y_2 + \lambda_{y_2}) = e^{\lambda_{y_2} D_{y_2}}\psi(y_1, y_2)\,,
\label{eq:pseudo-periodic}
\end{align}
under the translations $y_1\rightarrow y_1+ \lambda_{y_1}$ and $y_2\rightarrow y_2+\lambda_{y_2}$, respectively.
Here, $\chi_1$ and $\chi_2$ are rewritten by Wilson loops:
\begin{align}
    \chi_1 &= \lambda_{y_1} \oint_{\gamma_1} A(z) = \frac{\lambda_{y_1}\pi M}{{\rm Im}(\tau)}{\rm Im}(z) =\lambda_{y_1}  \pi M y_2\,,
    \nonumber\\
    \chi_2 &=  \lambda_{y_2} \oint_{\gamma_2} A(z) = \frac{\lambda_{y_2} \pi M}{{\rm Im}(\tau)}{\rm Im}(\Bar{\tau}z) = -\lambda_{y_2} \pi M y_1\,.    
\end{align}

The discrete translations $e^{\frac{n_{y_1}}{M}D_{y_1}}$ and $e^{\frac{n_{y_2}}{M}D_{y_2}}$ with $n_{y_1}, n_{y_2} = 0,1,...,M-1$ act on the wave functions of chiral zero modes:
\begin{align}
    Z^{n_{y_1}}\,&:\,e^{\frac{n_{y_1}}{M}D_{y_1}} \psi^{j,M}= e^{2\pi i\frac{j n_{y_1}}{M}} \psi^{j,M},
    \nonumber\\
    C^{n_{y_2}}\,&:\,e^{\frac{n_{y_2}}{M}D_{y_2}} \psi^{j,M}= \psi^{j+n_{y_2},M}.
    \label{eq:ZnxCny}
\end{align}
These zero modes may correspond to generations of quarks and leptons.
Thus, the discrete translations correspond to 
a flavor symmetry in 4D low-energy effective field theory.
They satisfy 
\begin{align}
    CZ= e^{\frac{2\pi i}{M}} ZC,
\label{eq:ZC}
\end{align}
for $n_{y_1} = n_{y_2} =1$ and generate the non-Abelian flavor symmetry $H_M\simeq (\mathbb{Z}_M \times \mathbb{Z}_M^{(Z)}) \rtimes \mathbb{Z}_M^{(C)}$, where 
$\mathbb{Z}_M^{(Z)}$ and $\mathbb{Z}_M^{(C)}$ respectively correspond to 
$Z$ and $C$, and the other $\mathbb{Z}_M$ corresponds to 
$e^{\frac{2\pi i}{M}}{\rm diag}(1,1,\cdots)$ \cite{Abe:2009vi,Berasaluce-Gonzalez:2012abm,Marchesano:2013ega}.

Next, we give a brief review of the coupling selection rule.
In order to study 3-point couplings, we extend the previous D-brane system to three stacks of D-branes $a$, $b$ and $c$, and they have $U(3)$ gauge symmetry if magnetic fluxes vanish. 
Let us introduce the following magnetic flux:
\begin{align}
    F = \frac{\pi i}{{\rm Im}\,(\tau)} (m_a X_a + m_b X_b+m_cX_c)
    dz \wedge d\Bar{z},
\end{align}
with 
\begin{align}
        X_a =
        \begin{pmatrix}
        1 & 0 & 0\\
        0 & 0 & 0\\
        0 & 0 & 0\\ 
    \end{pmatrix}
    ,\qquad
        X_b =
        \begin{pmatrix}
        0 & 0 & 0\\
        0 & 1 & 0\\
        0 & 0 & 0\\
    \end{pmatrix}
    ,\qquad
        X_c =
        \begin{pmatrix}
        0 & 0 & 0\\
        0 & 0 & 0\\
        0 & 0 & 1\\
    \end{pmatrix}.
\end{align}
This magnetic flux background breaks $U(3)$ to $U(1)_a\times U(1)_b \times U(1)_c$.
We consider the $ab$ matter sector, $bc$ matter sector, and $ac$ matter sector, 
and they have the charges, $(1,-1,0)$, $(0,1,-1)$, and $(1,0,-1)$, 
respectively under $U(1)_a\times U(1)_b \times U(1)_c$.
We denote their wave functions, $\psi^{i,M_{ab}}_{ab}$, 
$\psi^{j,M_{bc}}_{bc}$, $\psi^{k,M_{ac}}_{ac}$ by replacing 
$M$ with $M_{ab}$, $M_{bc}$, and $M_{ac}$ in Eq.~(\ref{eq:wf}), where 
$M_{ab}=m_a-m_b$, $M_{bc}=m_b-m_c$, $M_{ac}=m_a-m_c$. 
We set all of them to be positive, i.e., $M_{ab}, M_{bc}, M_{ac} > 0$.
Obviously, we find $M_{ac} = M_{ab}+M_{bc}$.
One of these three sectors must be a bosonic mode, but 
the bosonic mode has the same wave function as the fermionic mode 
in this background.
Hence, we adopt the same notation for wave functions of both bosonic and fermionic modes.

The 3-point coupling $y_{ijk}$ in 4D low energy effective field theory is obtained by the following overlap integration of wave functions:
\begin{align}
    y_{ijk}=g_3\int d^2z \psi^{i,M_{ab}}_{ab} \psi^{j,M_{bc}}_{bc} (\psi^{k,M_{ac}}_{ac})^*\,,
\end{align}
where $g_3$ denotes the coupling in higher dimensions.
The products of wave functions satisfy the following relation \cite{Cremades:2004wa}:
\begin{align}\label{eq:wf-product}
 \psi^{i,M_{ab}}_{ab} \psi^{j,M_{bc}}_{bc} = \sum_k c_{ijk}\psi^{k,M_{ac}}_{ac}\,,   
\end{align}
where
\begin{align}
    c_{ijk}&= {\cal A}^{-1/2} (2 {\rm Im}\,(\tau))^{1/4}\frac{M_{ab}M_{bc}}{M_{ac}} \notag \\ 
    &\times \sum_{m \in \mathbb{Z}_{M_{ac}}} \delta_{k,i+j+M_{ab}m}
       ~ \vartheta
\begin{bmatrix}
\frac{M_{bc}i-M_{ab}j+M_{ab}M_{bc}m}{M_{ab}M_{bc}M_{ac}}\\
0
\end{bmatrix}
(0, M_{ab}M_{bc}M_{ac}\tau)\,.
\end{align}
By use of this relation, the 3-point coupling in 4D effective field theory 
is written by $y_{ijk}=g_3c_{ijk}$.
The numbers in the Kronecker delta $\delta_{k,i+j+M_{ab}m}$ with $m \in \mathbb{Z}_{M_{ac}}$ is defined modulo $M_{ac}$.
That provides us with 
the coupling selection rule in 4D low energy effective field theory as 
\begin{align}
    i+j-k=M_{ac}\ell -M_{ab}m\,, 
\end{align}
where $m \in \mathbb{Z}_{M_{ac}}$ and $\ell \in \mathbb{Z}_{M_{ab}}$.
When ${\rm gcd}(M_{ab},M_{bc},M_{ac})=g$, 
the above condition becomes 
\begin{align}
    i+j-k=0 \qquad ({\rm mod}~ g)\,.
\end{align}
That implies that each matter sector has the symmetry $H_M\simeq (\mathbb{Z}_M \times \mathbb{Z}_M) \rtimes \mathbb{Z}_M$ depending on the magnetic flux $M=M_{ab}, M_{bc}$, or $M_{ac}$, but 
the 3-point coupling terms have only the common subsymmetry, i.e., 
$H_M\simeq (\mathbb{Z}_g \times \mathbb{Z}_g) \rtimes \mathbb{Z}_g$.
When $M_{ab}, M_{bc}$, and $M_{ac}$ are co-prime each other and  $g=1$, 
the result is trivial.

Similarly, we can calculate the 4-point coupling in 4D effective field theory by 
the following integration:
\begin{align}
    y_{ijk\ell}=g_4\int d^2z \psi^{i,M_1} \psi^{j,M_2}
    \psi^{k,M_3}(\psi^{\ell,M_4})^*\,,
\end{align}
where $g_4$ is the coupling in higher dimensions.
By use of the above relation (\ref{eq:wf-product}), we can write \cite{Abe:2009dr}
\begin{align}
    y_{ijk \ell} = g_4 \sum_m c_{ijm}c_{mk \ell}\,.
\end{align}
Furthermore, generic $n$-point couplings are also written by 
products of $c_{ijk}$.
Therefore, the coupling selection rule of the 3-point coupling also  
controls higher-order couplings.

For simplicity, we have considered $U(1)_a \times U(1)_b \times U(1)_c$ theory.
It is straightforward to extend it to $U(N_a) \times U(N_b) \times U(N_c)$ theory and study the selection rule of their bi-fundamental matter fields.

\subsection{Operator formalism}
\label{sec:operator}

In this section, we reformulate the magnetized D-brane system in a quantum mechanical system, following Refs. \cite{Cremades:2004wa,Abe:2014noa}. 
Again, we focus on two stacks of magnetized D-brane system on $T^2$. 
The charged fermions, i.e., the bifundamental representation of $U(1)_a \times U(1)_b$ with charge $(1,-1)$, obey the following Schr\"odinger equation:
\begin{align}
     H \psi (\bm{y}) = E \psi(\bm{y})\,,
\end{align}
with the Hamiltonian 
\begin{align}
    H = (-i \nabla + \bm{A}(\bm{y}))^2\,.
\label{eq:H}
\end{align}
Here and in what follows, $\nabla := (\partial_{y_1}, \partial_{y_2})^{\rm T}$ and the $U(1)$ gauge field is given as $A_i = -\frac{1}{2}F_{ij}y_j$ as in Eq.~\eqref{eq:Agauge}. 
In the operator formalism, the coordinate $\hat{y}_i$ and its canonical conjugate $\hat{\bm{p}} : = -i \nabla$ satisfy the commutation relations:
\begin{align}
    [\hat{y}_i, \hat{p}_j] = i \delta_{ij} \qquad (i,j=1,2)\,,
\end{align}
and otherwise 0. 
By rewriting the wave function as $\psi(\bm{y}) = \langle \bm{y}| \psi \rangle$ 
in the operator formalism, the state $|\psi\rangle$ obeys
\begin{align}
    \hat{H}|\psi\rangle = E |\psi\rangle\,,
\end{align}
with
\begin{align}
    \hat{H}  = \left( \hat{\bm{p}} - \frac{1}{2}F\hat{\bm{y}}\right)^2\,. 
\end{align}
Note that the state $|\psi \rangle$ is restricted to satisfy the pseudo-periodic boundary condition \eqref{eq:pseudo-periodic}:
\begin{align}
    e^{i\hat{T}_a} |\psi \rangle = |\psi \rangle\,,
    \label{eq:BC1}
\end{align}
with
\begin{align}
    \hat{T}_a := \bm{e}_a^{\rm T}\left( \hat{\bm{p}} - \frac{1}{2}F^{\rm T}\hat{\bm{y}}\right)\,,
\end{align}
where $\bm{e}_a$ $(a=1,2)$ is the basis vector of torus. 

Let us rewrite the operators $\{\hat{y}_1,\hat{y}_2,\hat{p}_1,\hat{p}_2\}$ as
\begin{align}
    \hat{Y}_1 &= \frac{\sqrt{2}}{\omega} \left( \hat{p}_2 - \frac{1}{2}F_{21}\hat{y}_1\right),
    \quad
    \hat{P}_1 = \sqrt{2}  \left( \hat{p}_1 - \frac{1}{2}F_{12}\hat{y}_2\right),
    \quad
    \hat{Y}_2 = -\frac{1}{2\pi M} \hat{T}_2,
    \quad
    \hat{P}_2 = \hat{T}_1,
\label{eq:YPdef}
\end{align}
with $\omega = 2F_{12}$, satisfying the canonical commutation relations:
\begin{align}
    [\hat{Y}_i, \hat{P}_j] = i \delta_{ij} \quad (i,j=1,2),
\label{eq:comrelation}
\end{align}
and otherwise 0. 
Then, the system is described by an one-dimensional harmonic oscillator:
\begin{align}
    \hat{H} = \frac{1}{2} \hat{P}_1^2 + \frac{\omega^2}{2} \hat{Y}_1^2\,,
\label{eq:Harmonic}
\end{align}
with the constraints
\begin{align}
    e^{i\hat{P}_2}|\psi \rangle = |\psi \rangle,
    \qquad
    e^{2\pi i M \hat{Y}_2}|\psi \rangle = |\psi \rangle.
    \label{eq:constraints}
\end{align}
Remarkably, two operators $\{ \hat{P}_2, \hat{Y}_2 \}$ distinguish degenerate states, although they commute with the Hamiltonian, i.e., conserved quantities. 
Indeed, when we choose an eigenstate of $\hat{Y}_2$ such that $\hat{Y}_2 |Y_2\rangle =Y_2 |Y_2\rangle$, the second constraint in Eq.~\eqref{eq:constraints} is rewritten as $e^{2\pi i M \hat{Y}_2} |Y_2\rangle =e^{2\pi i M Y_2}|Y_2\rangle$. 
Furthermore, the first constraint in Eq.~\eqref{eq:constraints} gives
\begin{align}
    e^{ia \hat{P}_2}|Y_2 \rangle = |Y_2 -a \rangle\,, 
\label{eq:Y2}
\end{align}
with $a \in \mathbb{R}$, which can be derived by acting $e^{2\pi i M \hat{Y}_2}$ on $e^{ia \hat{P}_2}|Y_2\rangle$ together with \eqref{eq:comrelation}. 
Since Eqs.~\eqref{eq:comrelation} and \eqref{eq:Y2} lead to the periodic condition $|Y_2\rangle = |Y_2 -1 \rangle$, we arrive at the quantization condition for the coordinate:
\begin{align}
Y_2 = \frac{j}{M} \quad (j=0,1,...,|M|-1)\,.
\label{eq:Y2_quantized}
\end{align}
Since the eigenvalue of $\hat{H}$ is determined by an ordinary one-dimensional harmonic oscillator \eqref{eq:Harmonic}:
\begin{align}
    E_n = \omega \left( n + \frac{1}{2}\right)\,,
\end{align}
the eigenstate of $\hat{H}$ is labeled by the following 
orthonormal basis:
\begin{align}
    \hat{H} |\psi\rangle_{T^2}^{j,M} = E_n  |\psi\rangle_{T^2}^{j,M}\,,
\end{align}
with $n=0,1,\cdots$, and
\begin{align}
    |\psi\rangle_{T^2}^{j,M} := \left| n, \frac{j}{M}\right\rangle\,.
\end{align}
Note that the following discussion is valid in every energy eigenstate. 
The above basis is generated by $\hat{P}_2$ which describes the translation along $\hat{Y}_2$ direction:
\begin{align}
    |\psi\rangle_{T^2}^{j,M} = e^{-i \frac{j}{M}\hat{P}_2}\left| n, 0\right\rangle\,.
\label{eq:Pgenerate}
\end{align}
So far, we have focused on the eigenvalues of two-dimensional Laplace operator \eqref{eq:H}, but the mass squared of the spinor and two-dimensional vector are given by $m_n^2 = \omega n$ and $m_n^2 = \omega (n-1/2)$, respectively.

\section{Non-invertible and flavor symmetries on $T^2/\mathbb{Z}_2$}
\label{sec:3}

In Sec. \ref{sec:translationsZN}, we first discuss the translation on $T^2/\mathbb{Z}_N$ which is restricted by the $\mathbb{Z}_N$ twist. 
Next, we define the non-invertible symmetry on $T^2/\mathbb{Z}_2$ by gauging the discrete group $\mathbb{Z}_2$ in Sec. \ref{sec:Z2}. 
Finally, we discuss the selection rule on $T^2/\mathbb{Z}_2$ under non-invertible flavor symmetries in Sec. \ref{sec:selectionrules}.

\subsection{Translations on $T^2/\mathbb{Z}_N$}
\label{sec:translationsZN}

We begin with the unorbifolded theory on the two-dimensional torus, which corresponds to the $c=2$ worldsheet conformal field theory. 
It was known that a topological operator ${\cal O}_{\delta X}$ inducing a translation on the sigma model fields $X^\mu \rightarrow X^\mu + \delta X^\mu$ is invertible on the $T^{2}$ torus. 
In toroidal orbifolds, the operator ${\cal O}_{\delta X}$ will not 
be preserved under the action of an orbifold group $\Gamma$ of isometries of the underlying $T^{2}$ torus, where ${\cal O}_{\delta X}$ transforms as ${\cal O}_{\delta X} \rightarrow g\,{\cal O}_{\delta X}\,g^{-1}$ with $g \in \Gamma$. 
Hence, after gauging the discrete group $\Gamma$, ${\cal O}_{\delta X}$ itself is not a gauge invariant quantity, but a sum of this operator over orbits $[\delta X]$ of the orbifold action $\oplus_{\delta X \in [\delta X]} {\cal O}_{\delta X}$ becomes a gauge invariant topological operator \cite{Heckman:2024obe}.\footnote{See, Ref.~\cite{Thorngren:2021yso}, how some invertible symmetries become non-invertible symmetries after gauging a discrete symmetry.}

In the magnetized extra dimensions, the system is described by one-dimensional quantum mechanics. When we gauge the orbifold group $\Gamma$, the operators on the torus are not gauge invariant operators on its orbifold. 
To specify the orbifold group, let us consider the two-dimensional rotation around the origin $y_1=y_2 =0$:
\begin{align}
    \bm{y} \rightarrow R_\theta \bm{y},
\end{align}
with 
\begin{align}
    R_\theta=
    \begin{pmatrix}
        \cos \theta & -\sin\theta\\
        \sin \theta & \cos \theta
    \end{pmatrix}
    ,
\end{align}
corresponding to the unitary operator $\hat{U}_\theta$:
\begin{align}
    \hat{\bm{y}} &\rightarrow \hat{U}_\theta \hat{\bm{y}} \hat{U}_\theta^\dagger = R_\theta \hat{\bm{y}}\,,
    \nonumber\\
    \hat{\bm{p}} & \rightarrow \hat{U}_\theta \hat{\bm{p}} \hat{U}_\theta^\dagger= R_\theta \hat{\bm{p}}\,.
\end{align}
Note that the Hamiltonian \eqref{eq:Harmonic} is invariant under the rotation $\hat{H} \rightarrow \hat{U}_\theta \hat{H} \hat{U}_\theta^\dagger$ due to $[R_\theta, F] =0$. 
However, two operators $\hat{T}_a$ imposing  the boundary conditions \eqref{eq:BC1} non-trivially transform as
\begin{align}
    \hat{T}_a \rightarrow \hat{U}_\theta \hat{T}_a \hat{U}_\theta^\dagger 
    = (R_\theta^{\rm T} \bm{e}_a )^{\rm T} \left( \hat{\bm{p}} - \frac{1}{2}F^{\rm T}\hat{\bm{y}}\right)\,,
\end{align}
with $a=1,2$. Following Ref. \cite{Abe:2014noa}, we choose the basis vector of the torus as
\begin{align}
    \mathbb{Z}_2\,&:\,(R_{\theta=\pi})^{\rm T} \bm{e}_1 = -   \bm{e}_1\,,
    \qquad \qquad \qquad(R_{\theta=\pi})^{\rm T} \bm{e}_2 = -   \bm{e}_2\,,
    \nonumber\\
    \mathbb{Z}_3\,&:\,(R_{\theta=2\pi/3})^{\rm T} \bm{e}_1 = -\bm{e}_1 -\bm{e}_2\,,
    \qquad (R_{\theta=2\pi/3})^{\rm T} \bm{e}_2 = \bm{e}_1\,,
    \nonumber\\
    \mathbb{Z}_4\,&:\,(R_{\theta=\pi/2})^{\rm T} \bm{e}_1 = -\bm{e}_2\,,
    \qquad (R_{\theta=\pi/2})^{\rm T} \bm{e}_2 = \bm{e}_1\,,
    \nonumber\\
    \mathbb{Z}_6\,&:\,(R_{\theta=\pi/3})^{\rm T} \bm{e}_1 = \bm{e}_1 -\bm{e}_2\,,
    \qquad (R_{\theta=\pi/3})^{\rm T} \bm{e}_2 = \bm{e}_1\,.
\end{align}
Hence, the discrete rotations $R_{\theta=2\pi/N}$ on $T^2/\mathbb{Z}_N$ induce the same transformation for $\hat{T}_a$ as
\begin{align}
    \mathbb{Z}_2\,&:\,\hat{T}_1 \rightarrow  -  \hat{T}_1\,,
    \qquad \hat{T}_2 \rightarrow  -  \hat{T}_2\,,
    \nonumber\\
    \mathbb{Z}_3\,&:\,\hat{T}_1 \rightarrow  -  \hat{T}_1 -  \hat{T}_2\,,
    \qquad \hat{T}_2 \rightarrow  \hat{T}_1\,,
    \nonumber\\
    \mathbb{Z}_4\,&:\,\hat{T}_1 \rightarrow  -  \hat{T}_2\,,
    \qquad \hat{T}_2 \rightarrow  \hat{T}_1\,,
    \nonumber\\
    \mathbb{Z}_6\,&:\,\hat{T}_1 \rightarrow  \hat{T}_1 -  \hat{T}_2\,,
    \qquad \hat{T}_2 \rightarrow  \hat{T}_1\,.
\label{eq:Z2Ta}
\end{align}
According to them, the operators $\{\hat{Y}_2, \hat{P}_2\}$ in Eq.~\eqref{eq:YPdef} transform under the discrete rotations $R_{\theta=2\pi/N}$. 
Hence, the gauging of $\mathbb{Z}_N$ restricts the form of $\{\hat{Y}_2, \hat{P}_2\}$ in $\mathbb{Z}_N$-invariant way, as will be discussed later.

\subsection{Non-invertible flavor symmetries}
\label{sec:Z2}

In this section, we focus on $T^2/\mathbb{Z}_2$ orbifold. 
Let us gauge the discrete group $\mathbb{Z}_2$, which corresponds to turning on the Kaluza-Klein gauge bosons associated with the isometries of the toroidal background. Note that the action of B-field on chiral zero modes is the trivial one. 
Then, the $\mathbb{Z}_2$ transformation \eqref{eq:Z2Ta} constrains the form of operators on $T^2/\mathbb{Z}_2$ as
\begin{align}
\hat{U}_{Y}^{(\lambda_Y)} \equiv e^{i \lambda_Y \hat{T}_2} + e^{- i \lambda_Y \hat{T}_2}\,,
\qquad
\hat{U}_{P}^{(\lambda_P)} \equiv e^{i\lambda_P \hat{T}_1} + e^{- i\lambda_P \hat{T}_1}\,,
\end{align}
where $\lambda_Y, \lambda_P \in \mathbb{R}$ is specified later, and 
$\{\hat{T}_1, \hat{T}_2\}$ respectively correspond to the canonical conjugate operators $\{\hat{P}_2, \hat{Y}_2\}$ as in Eq.~\eqref{eq:YPdef}. 
We find that they satisfy the fusion rule:
\begin{align}\label{eq:fusion-alg}
    \hat{U}_i^{(\lambda)}\hat{U}_i^{(\lambda^\prime)} &=\hat{U}_i^{(\lambda + \lambda^\prime)} + \hat{U}_i^{(\lambda - \lambda^\prime)}\,,
\end{align}
with $i=Y, P$. 
By using the Baker-Campbell-Hausdorff formula for $\hat{U}_Y$ and $\hat{U}_P$, 
we find
\small
\begin{align}
    \hat{U}_{Y}^{(\lambda_Y)}\hat{U}_{P}^{(\lambda_P)}
   &= \left(e^{i \lambda_Y \hat{T}_2 + i \lambda_P \hat{T}_1} + e^{-i  \lambda_Y \hat{T}_2 - i \lambda_P \hat{T}_1}\right) e^{-\pi i M \lambda_Y \lambda_P} 
    \nonumber\\
    &+ \left(e^{i \lambda_Y \hat{T}_2 - i \lambda_P \hat{T}_1} + e^{-i \lambda_Y \hat{T}_2 + i \lambda_P \hat{T}_1}\right) e^{\pi i M \lambda_Y \lambda_P} \,, 
    \nonumber\\
    \hat{U}_{P}^{(\lambda_P)}\hat{U}_{Y}^{(\lambda_Y)}
    &=\left(e^{i \lambda_Y \hat{T}_2 + i \lambda_P \hat{T}_1} + e^{-i  \lambda_Y \hat{T}_2 - i \lambda_P \hat{T}_1}\right) e^{\pi i M \lambda_Y \lambda_P} 
    \nonumber\\
    &+ \left(e^{i \lambda_Y \hat{T}_2 - i \lambda_P \hat{T}_1} + e^{-i \lambda_Y \hat{T}_2 + i \lambda_P \hat{T}_1}\right) e^{-\pi i M \lambda_Y \lambda_P} \,, 
\end{align}
\normalsize
with $[\hat{T}_2, \hat{T}_1] =2\pi i M$. 
Then, we arrive at the following algebraic relation:
\begin{align}
   \label{eq:comutation} 
   \hat{U}_{Y}^{(\lambda_Y)} \hat{U}_{P}^{(\lambda_P)} =e^{i\rho} \hat{U}_{P}^{(\lambda_P)} \hat{U}_{Y}^{(\lambda_Y)}\, ,
\end{align}
which is an orbifold version of  Eq.~(\ref{eq:ZC}). 
Since the coordinate $Y_2$ is now quantized as in Eq.~\eqref{eq:Y2_quantized} and $\hat{P}_2$ corresponds to the translation along $\hat{Y}$ direction as shown in Eq.~\eqref{eq:Pgenerate}, we choose\footnote{It corresponds to the discrete translations $e^{\frac{n_{y_1}}{M}D_{y_1}}$ and $e^{\frac{n_{y_2}}{M}D_{y_2}}$ leading to Eq.~\eqref{eq:ZnxCny}.}
\begin{align}
    \lambda_Y = \frac{n_Y}{M},\qquad
    \lambda_P = \frac{n_P}{M},
\end{align}
which gives rise to two possibilities:
\begin{align}
    &e^{i \rho}=1 \qquad ~~{\rm for}\,\,n_Y n_P = M n\,,
\nonumber\\
    &e^{i\rho}=-1 \qquad {\rm for}\,\,n_Y n_P = \frac{M}{2}\left( n+1\right)\,,
\end{align}
with $n\in \mathbb{Z}$. In the former case, one can perform the two Abelian transformations associated with $\hat{U}_Y$ and $\hat{U}_P$ simultaneously, but the latter case leads to the non-Abelian symmetry, as will be discussed below.

One of trivial representations of the fusion rule \eqref{eq:fusion-alg} 
corresponds to $\hat{U}_{Y}=0$ for all.
It also satisfies \eqref{eq:comutation}.
One has a similar trivial representation for $\hat{U}_{P}$, i.e. $\hat{U}_{P}=0$ for all.
Another trivial one corresponds to $\hat{U}_{Y}=2 \times \mathbb{I}_{n}$ for all, 
where $\mathbb{I}_{n}$ is the $(n \times n)$ unit matrix.
On this representation, we must have $e^{i\rho}=1$.
Also, we have a similar representation for $\hat{U}_{P}$, i.e. $\hat{U}_{P} = 2 \times \mathbb{I}_n$ for all. 
We study other representations by wave functions on orbifolds with magnetic fluxes.

On $T^2/\mathbb{Z}_2$ orbifold, $\mathbb{Z}_2$-even and -odd states are known as \cite{Abe:2008fi,Abe:2013bca,Abe:2014noa}
\begin{align}\label{eq:wf-Z2}
    |\psi\rangle_{\mathbb{Z}_2,+}^{j,M} &= \frac{1}{2}\left(  |\psi\rangle_{T^2}^{j,M} +  |\psi\rangle_{T^2}^{-j,M}  \right)\,,
    \nonumber\\
    |\psi\rangle_{\mathbb{Z}_2,-}^{j,M} &= \frac{1}{2}\left(  |\psi\rangle_{T^2}^{j,M} -  |\psi\rangle_{T^2}^{-j,M}  \right)\,.
\end{align} 
Let act the non-invertible transformations 
on the $\mathbb{Z}_2$-even states:
\begin{align}\label{eq:UxUy-wf}
    \hat{U}_{Y}^{(n_Y/M)}\,:\, \left(e^{i \frac{n_Y}{M} \hat{T}_2} + e^{- i\frac{n_Y}{M} \hat{T}_2}\right) |\psi\rangle_{\mathbb{Z}_2,+}^{j,M}
    &= 2\cos\left(2\pi \frac{jn_Y}{M} \right)|\psi\rangle_{\mathbb{Z}_2,+}^{j,M}\,,
    \nonumber\\
    \hat{U}_{P}^{(n_P/M)}\,:\, \left(e^{i\frac{n_P}{M} \hat{T}_1} + e^{- i\frac{n_P}{M} \hat{T}_1}\right) |\psi\rangle_{+}^{j,M}
    &=  |\psi\rangle_{\mathbb{Z}_2,+}^{j+n_P,M} +  |\psi\rangle_{\mathbb{Z}_2,+}^{j-n_P,M}\,.
\end{align}

It is remarkable that the fusion rule (\ref{eq:fusion-alg}) and 
transformation behavior of wave functions under $U_{\hat{Y}}$ are the same as 
the fusion rule and transformation behavior of operators in 
conformal field theory on $S^1/\mathbb{Z}_2$ studied in Ref.~\cite{Kaidi:2024wio}.
Note that $\hat{U}_{Y}$ always has the factor $2$.
In what follows, we rescale $\tilde{\hat{U}}_{Y}=\hat{U}_{Y}/2$ as well as $\tilde{\hat{U}}_{P}= \hat{U}_{P}/2$.

\paragraph{$M=2$}\,\\

When $M=2$, all modes are $\mathbb{Z}_2$-even and they are written by
\begin{align}
        |\Psi\rangle_{\mathbb{Z}_2}^{M=2}:=
    \begin{pmatrix}
        |\psi\rangle_{\mathbb{Z}_2,+}^{0,2}\\
        |\psi\rangle_{\mathbb{Z}_2,+}^{1,2}\\
    \end{pmatrix}
    .
\end{align}

First, we study the case leading to $e^{i\rho}=-1$, which is 
realized by $n_Y = n_P =1$.
Under the non-invertible transformations with $n_Y = n_P =1$, they transform as
\begin{align}
        |\Psi\rangle_{\mathbb{Z}_2}^{M=2}
    &\rightarrow 
    \begin{pmatrix}
        1 & 0\\
        0 & -1
    \end{pmatrix}
        |\Psi\rangle_{\mathbb{Z}_2}^{M=2}
    ,
    \nonumber\\
        |\Psi\rangle_{\mathbb{Z}_2}^{M=2}
    &\rightarrow 
    \begin{pmatrix}
        0 & 1\\
        1 & 0
    \end{pmatrix}
        |\Psi\rangle_{\mathbb{Z}_2}^{M=2}
    .
\end{align}
Obviously, these representations satisfy Eq.~\eqref{eq:fusion-alg} by the following identification:
\begin{align}
    &\hat{U}_{Y}^{(\lambda)}=\hat{U}_{Y}^{(\lambda')}=2\tilde{\hat{U}}_{Y} = 2\sigma_3\,, \quad 
    \hat{U}_{P}^{(\lambda)}= \hat{U}_{P}^{(\lambda')}=2\tilde{\hat{U}}_{P} = 2\sigma_1\,,
    \notag \\
    &\hat{U}_{Y}^{(\lambda+ \lambda')} = \hat{U}_{P}^{(\lambda+ \lambda')} =2\, \mathbb{I}_2\,,
\end{align}
where $\sigma_{1,3}$ denote the Pauli matrices.
Note that $\tilde{\hat{U}}_{Y}$ and $\tilde{\hat{U}}_{P}$ correspond to $Z$ and $C$ in Eq.~\eqref{eq:ZC}, respectively. 
Hence, the $\mathbb{Z}_2$-even modes belong to two-dimensional irreducible representation of 
$D_4\simeq (\mathbb{Z}_2\times \mathbb{Z}_2^{(\tilde{\hat{U}}_{Y})}) \rtimes \mathbb{Z}^{(\tilde{\hat{U}}_{P})}_2$, where $\mathbb{Z}_2^{(\tilde{\hat{U}}_{Y})}$ and $\mathbb{Z}^{(\tilde{\hat{U}}_{P})}_2$ respectively originate from $\tilde{\hat{U}}_{Y}$ and $\tilde{\hat{U}}_{P}$, and the other $\mathbb{Z}_2$ correspond to
\begin{align}
    \begin{pmatrix}
        -1 & 0\\
        0 & -1\\
    \end{pmatrix}.
\end{align}
These $\mathbb{Z}_2^{(\tilde{\hat{U}}_{Y})}$ and $\mathbb{Z}_2^{(\tilde{\hat{U}}_{P})}$ are related with $\mathbb{Z}_2^{(Z)}$ and 
$\mathbb{Z}_2^{(C)}$ in the previous section, respectively.

Next, we study the case leading to $e^{i \rho}=1$, which is 
realized by $n_Y=0$ or $n_P=0$.
That is, one of $\tilde{\hat{U}}_{Y}$ and $\tilde{\hat{U}}_{P}$ is trivial, i.e. 
$\tilde{\hat{U}}_{Y}=1$ or $\tilde{\hat{U}}_{P}=1$.
Then, the other operator,  $\mathbb{Z}_2^{\tilde{\hat{U}}_{Y}}$ or 
$\mathbb{Z}_2^{\tilde \hat{U}_{P}}$, remains.
It is a subgroup of $D_4$.
In what follows, we mainly discuss the case with 
$n_Y \neq 0$ and $n_P \neq 0$. That involves more.

\paragraph{$M=3$}\,\\

When $M=3$, there are two $\mathbb{Z}_2$-even modes:
\begin{align}
        |\Psi\rangle_{\mathbb{Z}_2}^{M=3}:=
    \begin{pmatrix}
        |\psi\rangle_{\mathbb{Z}_2,+}^{0,3}\\
        |\psi\rangle_{\mathbb{Z}_2,+}^{1,3}\\
    \end{pmatrix}
    \,.
\end{align}
When $M=3$, we have no solution for $e^{i \rho}=-1$.
A solution for $e^{i \rho}=1$ corresponds to $n_Y=0$ or $n_P=0$, where one of two is trivial.
Let us study other cases.
For example, under the non-invertible transformations with $n_Y= n_P=1$, the wave functions transform as
\begin{align}
    |\Psi\rangle_{\mathbb{Z}_2}^{M=3}
    &\rightarrow 
    \begin{pmatrix}
        1 & 0\\
        0 & \cos(\frac{2\pi}{3})
    \end{pmatrix}
    |\Psi\rangle_{\mathbb{Z}_2}^{M=3}
    ,
    \nonumber\\
    |\Psi\rangle_{\mathbb{Z}_2}^{M=3}
    &\rightarrow 
    \begin{pmatrix}
        0 & 1\\
        \frac{1}{2} & \frac{1}{2}
    \end{pmatrix}
    |\Psi\rangle_{\mathbb{Z}_2}^{M=3}
    .
\end{align}
When $n_Y=n_P=2$, we obtain the same transformations with $n_Y=n_P=1$. 
Obviously, they satisfy the fusion rule (\ref{eq:fusion-alg}):
\begin{align}
   & (2\tilde{\hat{U}}_i^{(\lambda)})(2\tilde{\hat{U}}_i^{(\lambda)})=2\tilde{\hat{U}}_i^{(\lambda)} + 
    2 \mathbb{I}_2\,,
\end{align}
with $i = Y, P$ and $\lambda= 1/3, 2/3$.

Note that the non-Abelian symmetry $H_3$ associated with Eq.~\eqref{eq:ZC} exists before the orbifold, but $\mathbb{Z}_3^{(Z)}$ and $\mathbb{Z}_3^{(C)}$ are broken 
by the $\mathbb{Z}_2$ orbifold twist. 
Similarly, when $M=$ odd, no "invertible" symmetry originating from $\mathbb{Z}_M^{(Z)}$ and $\mathbb{Z}_M^{(C)}$ remains.

\paragraph{$M=4$}\,\\

When $M=4$, the $\mathbb{Z}_2$-even modes are given by
\begin{align}
|\Psi\rangle_{\mathbb{Z}_2}^{M=4}:=
    \begin{pmatrix}
        |\psi\rangle_{\mathbb{Z}_2,+}^{0,4}\\
        |\psi\rangle_{\mathbb{Z}_2,+}^{1,4}\\
        |\psi\rangle_{\mathbb{Z}_2,+}^{2,4}\\
    \end{pmatrix}
    \,.
\end{align}

First, we study the case leading to $e^{i\rho} =-1$.
That can be realized by $(n_Y,n_P)=$ (even, odd) or (odd, even).
If $n_Y=$ odd, $\hat{U}_{Y}^{(n_Y)}$ includes vanishing eigenvalues in Eq.~\eqref{eq:UxUy-wf}. 
For example, when $(n_Y,n_P)=(1,2)$, the wave functions transform as
\begin{align}
    \tilde{\hat{U}}_{Y}^{(1/4)}\,&:\,        
    |\Psi\rangle_{\mathbb{Z}_2}^{M=4}
    \rightarrow 
        \begin{pmatrix}
        1 & 0 & 0\\
        0 & 0 & 0\\
        0 & 0 & -1 \\
    \end{pmatrix}
    |\Psi\rangle_{\mathbb{Z}_2}^{M=4}
    \,,
    \nonumber\\
   \tilde{\hat{U}}_{P}^{(1/2)}\,&:\,       
    |\Psi\rangle_{\mathbb{Z}_2}^{M=4}
    \rightarrow 
        \begin{pmatrix}
        0  & 0 & 1\\
        0  & 1 & 0\\
        1 & 0 & 0\\
    \end{pmatrix}
    |\Psi\rangle_{\mathbb{Z}_2}^{M=4}
    \,.
\end{align}
By changing the basis, these transformation matrices can be written by
\begin{align}
    \begin{pmatrix}
        \sigma_3 & 0 \\
        0 & 0\\
    \end{pmatrix},
    \qquad 
    \begin{pmatrix}
        \sigma_1 & 0 \\
        0 & 1
    \end{pmatrix},
\end{align}
which are the direct sum of the representation studied in $M=2$ and the 
trivial representation.
They are also represented by changing the basis as,
\begin{align}
    \begin{pmatrix}
        \sigma_1 & 0 \\
        0 & 0\\
    \end{pmatrix},
    \qquad 
    \begin{pmatrix}
        \sigma_3 & 0 \\
        0 & 1
    \end{pmatrix}.
\end{align}
Similarly, if $n_P=$ odd, we find ${\rm det}\,\hat{U}_P^{(n_P/M)}=0$, i.e., 
$\hat{U}_P^{(n_P/M)}$ includes vanishing eigenvalues. 
Through exchanging $\tilde{\hat{U}}_Y$ and $\tilde{\hat{U}}_P$, 
the representation matrix of chiral zero modes are represented by the above matrices. 
Hence, the case with $e^{i\rho} =-1$ includes the trivial representation.

Next, we study the case leading to $e^{i\rho} =1$.
That can be realized by $n_Y=n_P=2$. 
Under the non-invertible transformations with $n_Y=n_P=2$, they transform as
\begin{align}
    \tilde{\hat{U}}_Y^{(1/2)}\,&:\,        
    |\Psi\rangle_{\mathbb{Z}_2}^{M=4}
    \rightarrow 
        \begin{pmatrix}
        1 & 0 & 0\\
        0 & -1 & 0\\
        0 & 0 & 1 \\
    \end{pmatrix}
    |\Psi\rangle_{\mathbb{Z}_2}^{M=4}
    \,,
    \nonumber\\
   \tilde{\hat{U}}_P^{(1/2)}\,&:\,       
    |\Psi\rangle_{\mathbb{Z}_2}^{M=4}
    \rightarrow 
        \begin{pmatrix}
        0  & 0 & 1\\
        0  & 1 & 0\\
        1 & 0 & 0\\
    \end{pmatrix}
    |\Psi\rangle_{\mathbb{Z}_2}^{M=4}
    \,.
\end{align}
Hence, $\mathbb{Z}_2^{(\tilde{\hat{U}}_Y)}$ and $\mathbb{Z}^{(\tilde{\hat{U}}_P)}_2$ appear.
The pair $(|\psi\rangle_{T^2}^{0,4},|\psi\rangle_{T^2}^{2,4})$ can be decomposed into two singlets, 
and they can be denoted by $1_{++}$ and $1_{+-}$,
where two subscripts mean $\mathbb{Z}_2$ even or odd for 
$\mathbb{Z}_2^{(\tilde{\hat{U}}_Y)}$ and $\mathbb{Z}^{(\tilde{\hat{U}}_P)}_2$.
The other mode corresponds to $1_{-+}$. 
Note that $\mathbb{Z}_2^{(\tilde{\hat{U}}_Y)} \times \mathbb{Z}^{(\tilde{\hat{U}}_P)}_2$ is a subgroup of $D_4$, and $D_4$ has four singlets, $1_{++}, 1_{+-}, 1_{-+}, 1_{--}$.
(See e.g., Refs.~\cite{Ishimori:2010au,Kobayashi:2022moq}.)
The mode corresponding to $1_{--}$ is the $\mathbb{Z}_2$-odd mode:
\begin{align}
       |\psi\rangle_{T^2}^{1,4} - |\psi\rangle_{T^2}^{3,4}\,.
\end{align}

Similarly, we can discuss the case with $M=$ even.
When $M=2k$ and $n_Y=n_P=k$, eigenvalues do not include $0$.
We concentrate on this case.

\paragraph{$M=2k$}\,\\

When $M=2k$ with $k\in \mathbb{Z}$, the $\mathbb{Z}_2$-even modes are given by
\begin{align}
|\Psi\rangle_{\mathbb{Z}_2}^{M=2k}:=
    \begin{pmatrix}
        |\psi\rangle_{+}^{0,2k}\\
        |\psi\rangle_{+}^{j,2k}\\
        |\psi\rangle_{+}^{k,2k}\\
    \end{pmatrix}
    \,,
\end{align}
with $j=1,2,...,k-1$. 
Under the non-invertible transformations with $n_Y=n_P=k$, they transform as
\begin{align}
   \tilde{\hat{U}}_Y^{(k/M)}\,&:\,        
    |\Psi\rangle_{\mathbb{Z}_2}^{M=2k}
    \rightarrow 
        \begin{pmatrix}
        1 & 0 & \cdots & \cdots & \cdots &  0 & 0\\
        0 & -1 & \cdots & \cdots & \cdots & 0 & 0\\
        \vdots & \vdots &  &  &  & \vdots & \vdots \\
        0 & 0 & \cdots & \cos (\pi j) & \cdots & 0 & 0\\
        \vdots & \vdots &  &  &  & \vdots & \vdots \\
        0 & 0 & \cdots & \cdots & \cdots & -\cos(\pi k) & 0 \\
        0 & 0 & \cdots & \cdots & \cdots & 0 & \cos(\pi k) \\
    \end{pmatrix}
|\Psi\rangle_{\mathbb{Z}_2}^{M=2k}
    \,,
    \nonumber\\
    \tilde{\hat{U}}_P^{(k/M)}\,&:\,       
    |\Psi\rangle_{\mathbb{Z}_2}^{M=2k}
    \rightarrow 
        \begin{pmatrix}
        0  & 0 & \cdots &  0 & 1\\
        0  & 0 & \cdots &  1 & 0\\
        \vdots & \vdots & & \vdots & \vdots \\
        0 & 1 & \cdots  & 0 & 0 \\
        1 & 0 & \cdots & 0 & 0\\
    \end{pmatrix}
|\Psi\rangle_{\mathbb{Z}_2}^{M=2k}
    \,.
\end{align}

\paragraph{$k=$ even}\,\\

When $k$ is even, the pairs with the same $\mathbb{Z}_2^{(\tilde{\hat{U}}_Y)}$ charges transform each other by $\tilde{\hat{U}}_P$.
Thus, they correspond to singlets, $1_{+\pm}$ and $1_{-\pm}$ of $D_4$.

\paragraph{$k=$ odd}\,\\

When $k$ is odd, the pairs with different $\mathbb{Z}_2^{(\tilde{\hat{U}}_Y)}$ charges transform each other by $\tilde{\hat{U}}_P$.
Thus, they correspond to $D_4$ doublets.

\subsection{Coupling selection rules}
\label{sec:selectionrules}

The wave functions on $T^2/\mathbb{Z}_2$ are constructed by 
ones on $T^2$.
Hence, the coupling selection rule on $T^2/\mathbb{Z}_2$ is derived from 
one on $T^2$.
For example, the allowed 3-point couplings $y_{ijk}$ of 
$U(1)_a \times U(1)_b \times U(1)_c$ theory on $T^2$ must satisfy 
$k=i+j+M_{ab}m$ with $m =M_{ac}n$, where $n$ is integer. 
The wave function $\psi^{j,M}$ on $T^2$ is combined with $\psi^{M-j,M}$ as Eq.~(\ref{eq:wf-Z2}) 
on $T^2/\mathbb{Z}_2$.
The coupling selection rule can be written as
\begin{align}\label{eq:selection-orbi}
    [k]=[i]+[j] \quad ({\rm mod}\,M_{ab}M_{ac})\,,
\end{align}
where $[i]$ denotes $i$ or $M_{ab} -i$, and $[j]$ and $[k]$ have a similar meaning.

Now let us study coupling selection rules among matter fields with $M=$even.
Those matter fields have the common flavor symmetry $D_4$ when $n_Y=n_P=M/2$.
Their coupling selection rules are consistent with $D_4$.
For example, we consider the 3-point coupling among matter fields with 
$M_{ab}=M_{bc}=2$ and $M_{ac}=4$.
The zero modes $\psi_{ab}^{i,2}$ and $\psi_{bc}^{j,2}$ with $i,j=0,1$ are $D_4$ doublets. 
Here and in what follows, we adopt the language of the wave function. 
Their tensor product is decomposed as 
\begin{align}
   &(\psi^{0,2}_{ab}\psi^{0,2}_{bc}+\psi^{1,2}_{ab}\psi^{1,2}_{bc})_{1_{++}} 
   +(\psi^{0,2}_{ab}\psi^{0,2}_{bc}-\psi^{1,2}_{ab}\psi^{1,2}_{bc})_{1_{+-}}  
   \notag \\
   &+(\psi^{0,2}_{ab}\psi^{1,2}_{bc}+\psi^{0,2}_{ab}\psi^{1,2}_{bc})_{1_{-+}} 
   +(\psi^{0,2}_{ab}\psi^{1,2}_{bc}-\psi^{0,2}_{ab}\psi^{1,2}_{bc})_{1_{-+}}\,, 
\end{align}
where subscripts denote their representations under $D_4$.
They couple with $\psi^{k,4}_+$ when their $\mathbb{Z}_2\times \mathbb{Z}_2$ 
charges are the same.
These coupling selection rules are consistent with Eq.~(\ref{eq:selection-orbi}). 
Also, for other couplings including only $M=$even, the $D_4$ flavor symmetry is 
consistent with Eq.~(\ref{eq:selection-orbi}).

When the 3-point couplings include the modes with $M=$odd, 
invertible symmetry does not exist.
However, the coupling selection rule (\ref{eq:selection-orbi}) is non-trivial 
even when $M=$ odd.
There may be some "symmetry" behind this selection rule.
Here, we discuss briefly.
We use the model with $M_{ab}=M_{bc}=3$ and $M_{ac}=6$ as an illustrating example.
In this model, the coupling selection rule 
(\ref{eq:selection-orbi}) tells the following 
expansions of wave function products:
\begin{align}
\label{eq:selection-M=3-1}
&    \psi^{0,3}_+ \psi^{0,3}_+ \sim c_{000}\psi^{0,6}_+ +
    c_{003}\psi^{3,6}_+\,, \\
   \label{eq:selection-M=3-2} 
    &    \psi^{0,3}_+ \psi^{1,3}_+ \sim c_{011}\psi^{1,6}_+ +
    c_{012}\psi^{2,6}_+\,, \\
    \label{eq:selection-M=3-3}
    &    \psi^{1,3}_+ \psi^{1,3}_+ \sim c_{110}\psi^{0,6}_+ + c_{111}\psi^{1,6}_+ + c_{112}\psi^{2,6}_++ c_{113}\psi^{3,6}_+\,.
\end{align}
In what follows, we do not discuss explicit values of the coefficients $c_{ijk}$, because they may not be discussed by symmetries. 
The generator $\tilde{\hat U}_Y^{(\lambda)}$ is 
represented by $\tilde{\hat U}_Y^{(\lambda)}=1$ on 
$\psi^{0,3}_+$ and $\psi^{0,6}_+$  as well as 
$\psi^{3,6}_+$.
This behavior is consistent with Eq.~(\ref{eq:selection-M=3-1}).
In general, the generator $\tilde{\hat U}_Y^{(\lambda)}$ is represented on the wave function with $\psi^{0,M}_+$ by $\tilde{\hat{U}}_Y^{(\lambda)}=1$.

Next, we operate $\tilde{\hat U}_Y^{(1/3)}$ on both sides of 
Eq.~(\ref{eq:selection-M=3-2}).
Eigenvalues of both sides are the same, i.e. $\cos (2\pi/3)$.
Thus, Eq.~(\ref{eq:selection-M=3-2}) is also consistent with the non-invertible symmetry.
Thus, the tensor products and the coupling selection rule seem to be simple when the products include the states with $\tilde{\hat U}_Y^{(\lambda)}=1$.
However, such simple interpretation does not seem to be applicable in Eq.~(\ref{eq:selection-M=3-3}).\footnote{The selection rule in Eq.~(\ref{eq:selection-M=3-3}) could be understood by the modified $\mathbb{Z}_3$ symmetry, where the $\mathbb{Z}_3$ charges, 1 and 2 are identified by gauging the $\mathbb{Z}_2$ orbifold twist.}
We may need to use the fusion rule.
At any rate, we have found that both Eqs.~(\ref{eq:selection-M=3-1}) and (\ref{eq:selection-M=3-2}) are consistent with $\tilde{\hat U}_Y^{(\lambda)}$.
Note that their right-hand sides are related by $\tilde{\hat U}_P^{(1/3)}$, i.e.,
\begin{align}
    \tilde{\hat U}_P^{(1/3)} (\psi^{0,6}_+ + \psi^{3,6}) \sim 
    \psi^{1,6}_+ + \psi^{2,6},
\end{align}
up to coefficients.
That suggests that 
\begin{align}
    \tilde{\hat U}^{(1/3)}_P (\psi^{0,3}_+ \psi^{0,3}_+ )
    \sim \psi^{0,3}_+ \psi^{1,3}_+ + \psi^{1,3}_+ \psi^{0,3}_+.
\end{align}
That may be a rule for tensor products.
Then, the selection rules (\ref{eq:selection-M=3-1}) and (\ref{eq:selection-M=3-2}) become consistent with 
$\tilde{\hat U}_Y$ and $\tilde{\hat U}_P$.
Similarly, we can discuss Eq.~(\ref{eq:selection-M=3-3}) by multiplying Eq.~(\ref{eq:selection-M=3-2}) by 
$\tilde{\hat U}_P$.
Alternatively, Eq.~(\ref{eq:selection-M=3-3}) could be derived as follows.
We can write $\psi^{1,3}_+ \psi^{1,3}_+ \sim 
(\tilde{\hat U}^{(1/3)}_P\psi^{0,3}_+)(\tilde{\hat U}^{(1/3)}_P\psi^{0,3}_+)$.
Suppose that this term can be written by
$(\mathbb{I}+\tilde{\hat U}^{(1/3)}_P)\psi^{0,3}_+ \psi^{0,3}_+$ through the fusion rule, up to coefficients.
Then, we use of Eq.~(\ref{eq:selection-M=3-1}) so as to find 
\begin{align}
    (\mathbb{I}+\tilde{\hat U}^{(1/3)}_P)(\psi^{0,6}_+ +\psi^{3,6}_+) 
    \sim \psi^{0,6}_+ +\psi^{1,6}_+ +\psi^{2,6}_+ + \psi^{3,6}_+,
\end{align}
which corresponds to the right-hand side of Eq.~(\ref{eq:selection-M=3-3}).
Thus, the coupling selection rule may have a certain relation with the non-invertible symmetry 
generated by $\tilde{\hat U}_Y$ and $\tilde{\hat U}_P$.
To verify this statement, we have to understand the tensor products and the selection rules by the non-invertible symmetry 
in the framework of modular tensor category. 
At any rate, 
the non-invertible symmetry may restrict the structure 
of the selection rule. Detailed studies including other fluxes are important. We leave them as a future work.

\section{Non-invertible and flavor symmetries on $T^2/\mathbb{Z}_3$, $T^2/\mathbb{Z}_4$ and $T^2/\mathbb{Z}_6$}
\label{sec:4}

In this section, we extend the analysis of $\mathbb{Z}_2$ to $\mathbb{Z}_N$ orbifolds 
with $N=3,4,6$. 
By deriving the fusion rule on $\mathbb{Z}_N$ orbifolds in Sec. \ref{sec:ZNfusion}, we respectively show the representation of chiral zero modes on $T^2/\mathbb{Z}_3$, $T^2/\mathbb{Z}_4$ and $T^2/\mathbb{Z}_6$ in Secs. \ref{sec:Z3}, \ref{sec:Z4} and \ref{sec:Z6}.

\subsection{Fusion rule on $T^2/\mathbb{Z}_N$}
\label{sec:ZNfusion}

We discuss two stacks of magnetized D-branes on $T^2/\mathbb{Z}_N$ orbifolds with $N=3,4,6$. 
Under the $\mathbb{Z}_N$ action $\theta : z\rightarrow \omega z$ with $\omega =e^{\frac{2\pi i}{N}}$, the isometry of underlying $T^2$ is broken down since the complex structure of $T^2$ is fixed at $\tau = \omega$. 
Taking into account the discrete rotation on $T^2/\mathbb{Z}_N$ \eqref{eq:Z2Ta}, 
the following operators are $\mathbb{Z}_N$ invariant after gauging $\mathbb{Z}_N$:
\begin{align}
\hat{U}_{\mathbb{Z}_3}^{(\lambda_1, \lambda_2)} \equiv &\,\,e^{i\lambda_1 \hat{T}_1+i \lambda_2 \hat{T}_2}+ e^{i \lambda_1 \hat{T}_2 -i
\lambda_2(\hat{T}_1+\hat{T}_2)} + e^{-i\lambda_1 (\hat{T}_1+\hat{T}_2)  +i
\lambda_2 \hat{T}_1}\,,
\nonumber\\
\hat{U}_{\mathbb{Z}_4}^{(\lambda_1, \lambda_2)} \equiv  &\,\,e^{i\lambda_1 \hat{T}_1 +i\lambda_2 \hat{T}_2 } + e^{-i\lambda_1 \hat{T}_2 + i\lambda_2 \hat{T}_1} + e^{-i\lambda_1 \hat{T}_1 - i\lambda_2 \hat{T}_2}
+ e^{i\lambda_1 \hat{T}_2 - i\lambda_2 \hat{T}_1}\,,
\nonumber\\
\hat{U}_{\mathbb{Z}_6}^{(\lambda_1, \lambda_2)} \equiv  &\,\,e^{i\lambda_1 \hat{T}_1 +i\lambda_2 \hat{T}_2} + e^{i\lambda_1 (\hat{T}_1-\hat{T}_2) + i\lambda_2 \hat{T}_1} + e^{-i\lambda_1 \hat{T}_2 + i\lambda_2 (\hat{T}_1-\hat{T}_2)}
+ e^{-i\lambda_1 \hat{T}_1 - i\lambda_2 \hat{T}_2}
\nonumber\\
&+ e^{-i\lambda_1 (\hat{T}_1-\hat{T}_2) - i\lambda_2 \hat{T}_1}
+ e^{i\lambda_1 \hat{T}_2 - i\lambda_2 (\hat{T}_1 -\hat{T}_2)}\,.
\label{eq:UZN}
\end{align}
We find that they obey the following fusion rule:
\small
\begin{align}
\hat{U}_{\mathbb{Z}_3}^{(\lambda_1, \lambda_2)}\hat{U}_{\mathbb{Z}_3}^{(\lambda_3, \lambda_4)} 
&=e^{\pi i M (\lambda_2\lambda_3 - \lambda_1 \lambda_4)} \hat{U}_{\mathbb{Z}_3}^{(\lambda_1+\lambda_3, \lambda_2+\lambda_4)} 
+ e^{\pi i M (\lambda_1\lambda_3-\lambda_2\lambda_3 + \lambda_2 \lambda_4)} \hat{U}_{\mathbb{Z}_3}^{(-\lambda_1+\lambda_2-\lambda_4, -\lambda_1 +\lambda_3 -\lambda_4)}
\nonumber\\
&+ e^{\pi i M (\lambda_1\lambda_4-\lambda_2\lambda_4 - \lambda_1 \lambda_3)} \hat{U}_{\mathbb{Z}_3}^{(\lambda_1-\lambda_4, \lambda_2 +\lambda_3 -\lambda_4)}\,,
\nonumber\\
\hat{U}_{\mathbb{Z}_4}^{(\lambda_1, \lambda_2)}\hat{U}_{\mathbb{Z}_4}^{(\lambda_3, \lambda_4)} 
&=e^{\pi i M (\lambda_2\lambda_3 - \lambda_1 \lambda_4)} \hat{U}_{\mathbb{Z}_4}^{(\lambda_1+\lambda_3, \lambda_2+\lambda_4)} 
+ e^{-\pi i M (\lambda_1\lambda_3 + \lambda_2 \lambda_4)} \hat{U}_{\mathbb{Z}_4}^{(\lambda_1-\lambda_4, \lambda_2 +\lambda_3)}
\nonumber\\
&+ e^{\pi i M (\lambda_1\lambda_4-\lambda_2\lambda_3)} \hat{U}_{\mathbb{Z}_4}^{(\lambda_1-\lambda_3, \lambda_2 -\lambda_4)}
+ e^{\pi i M (\lambda_1\lambda_3 +\lambda_2\lambda_4)} \hat{U}_{\mathbb{Z}_4}^{(\lambda_1 +\lambda_4, \lambda_2 -\lambda_3)}\,,
\nonumber\\
\hat{U}_{\mathbb{Z}_6}^{(\lambda_1, \lambda_2)}U_{\mathbb{Z}_6}^{(\lambda_3, \lambda_4)} 
&=e^{\pi i M (\lambda_2\lambda_3 - \lambda_1 \lambda_4)} \hat{U}_{\mathbb{Z}_6}^{(\lambda_1+\lambda_3, \lambda_2+\lambda_4)} 
+ e^{-\pi i M (\lambda_1\lambda_3 +\lambda_1\lambda_4 + \lambda_2 \lambda_4)} \hat{U}_{\mathbb{Z}_6}^{(\lambda_1+\lambda_2+\lambda_3, -\lambda_1 +\lambda_4)}
\nonumber\\
&+ e^{-\pi i M (\lambda_1\lambda_3 +\lambda_2\lambda_3 + \lambda_2 \lambda_4)} \hat{U}_{\mathbb{Z}_6}^{(\lambda_2 +\lambda_3, -\lambda_1 -\lambda_2+\lambda_4)}
+ e^{\pi i M (\lambda_1\lambda_4- \lambda_2 \lambda_3)} \hat{U}_{\mathbb{Z}_6}^{(\lambda_3-\lambda_1, \lambda_4 -\lambda_2)}
\nonumber\\
&+ e^{\pi i M (\lambda_1\lambda_3 +\lambda_1\lambda_4 + \lambda_2 \lambda_4)} \hat{U}_{\mathbb{Z}_6}^{(\lambda_3 -\lambda_1-\lambda_2, \lambda_1 +\lambda_4)}
+ e^{\pi i M (\lambda_1\lambda_3 +\lambda_2\lambda_3 + \lambda_2 \lambda_4)} \hat{U}_{\mathbb{Z}_6}^{(\lambda_3-\lambda_2, \lambda_1 +\lambda_2 +\lambda_4)}\,.
\label{eq:fusionrulesZN}
\end{align}
\normalsize
Similar to $T^2/\mathbb{Z}_2$, a trivial representation corresponds to $\hat{U}^{(\lambda_1, \lambda_2)}_{\mathbb{Z}_N}=N\times \mathbb{I}$ or $0$ for all elements.

The chiral zero modes on $T^2/\mathbb{Z}_N$ are obtained as a linear combination of matter wave functions on $T^2$ \cite{Abe:2013bca,Abe:2014noa,Kobayashi:2017dyu}:
\begin{align}
|\psi\rangle_{\mathbb{Z}_N,\eta}^{j,M} &= \Hat{P}_{\mathbb{Z}_N,\eta}|\psi\rangle_{T^2}^{j,M}\,,
\label{eq:Pstate}
\end{align}
with $\hat{P}_{\mathbb{Z}_M,\eta}$ being the projection operator: 
\begin{align}
\hat{P}_{\mathbb{Z}_N,\eta} := 
\frac{1}{N}\left\{ 
\sum_{x=0}^{|M|-1} 
\Bar{\eta}^x(\hat{U}_{\mathbb{Z}_N})^x\right\}
\,.
\end{align}
Here, $\eta$ denotes the $\mathbb{Z}_N$ eigenvalues, i.e., $\eta=1,\omega, \omega^2,...,\omega^{N-1}$.

Let us act the non-invertible transformations \eqref{eq:UZN} on the state $|\psi\rangle_{\mathbb{Z}_N,\eta}^{j,M}$. Since $\hat{U}_{\mathbb{Z}_N}^{(\lambda_1, \lambda_2)}$ is invariant under the $\mathbb{Z}_N$-twist, it obeys
\begin{align}
\hat{U}_{\mathbb{Z}_N} \hat{U}_{\mathbb{Z}_N}^{(\lambda_1, \lambda_2)}\hat{U}_{\mathbb{Z}_N}^{-1} = \hat{U}_{\mathbb{Z}_N}^{(\lambda_1, \lambda_2)}\,.
\end{align}
Hence, we find that $\hat{U}_{\mathbb{Z}_N}^{(\lambda_1, \lambda_2)}$ and the projection operator $\hat{P}_{\mathbb{Z}_N,\eta}$ is commutable, i.e., 
\begin{align}
\hat{U}_{\mathbb{Z}_N}^{(\lambda_1, \lambda_2)}|\psi\rangle_{\mathbb{Z}_N,\eta}^{j,M} &=  \hat{P}_{\mathbb{Z}_N,\eta} \hat{U}_{\mathbb{Z}_N}^{(\lambda_1, \lambda_2)}|\psi\rangle_{T^2}^{j,M}\,.
\label{eq:UPcom}
\end{align}
In the following sections, we discuss the non-invertible transformation of chiral zero modes on $T^2/\mathbb{Z}_N$ orbifolds with $N=3,4,6$. 

\subsection{Non-invertible symmetries on $T^2/\mathbb{Z}_3$}
\label{sec:Z3}

It was known in Ref. \cite{Abe:2014noa} that the projected states on $T^2/\mathbb{Z}_3$ are given by
\begin{align}
|\psi\rangle_{\mathbb{Z}_3,\eta}^{j,M} &= \sum_{k=0}^{|M|-1} M_{jk}^{(\mathbb{Z}_N,\eta)}|\psi\rangle_{T^2}^{k,M} \,,     
\end{align}
with 
\begin{align}
M_{jk}^{(\mathbb{Z}_N,\eta)} =\frac{1}{N}\sum_{x=0}^{N-1}\bar{\eta}^x \sum_{k=0}^{|M|-1} D_{jk}^{\omega^x}\,,
\end{align}
where $\eta$ denotes the $\mathbb{Z}_3$ eigenvalues, i.e., $\eta=1,\omega, \omega^2$, and 
\begin{align}
    D_{k}^{j} &:= \frac{1}{\sqrt{M}}e^{-i \frac{\pi}{12} +i \frac{\pi k^2}{M} + 2\pi i \frac{jk}{M}}\,,
\nonumber\\
    (D^{-1})_{k}^{j} &:= \frac{1}{\sqrt{M}}e^{i \frac{\pi}{12} -i \frac{\pi j^2}{M} - 2\pi i \frac{jk}{M}}\,. 
\end{align}
Here and in what follows, we focus on the case with $M>0$. 
Note that in the language of wave function, the $\mathbb{Z}_3$-twisted mode function $\psi_{T^2}^{j,M}(\omega^xz)$ with $x=1,2$ can be expanded by $\psi_{T^2}^{j,M}(z)$~\cite{Abe:2014noa}\footnote{Note that $D_k^j$ corresponds to $\Tilde{D}_k^j$ in Ref. \cite{Kobayashi:2017dyu}.}
\begin{align}
    \psi_{T^2}^{j,M}(\omega z) &= \sum_{k=0}^{|M|-1} D^j_{k} \psi_{T^2}^{k,M}(z)\,,
    \nonumber\\
    \psi_{T^2}^{j,M}(\omega^2 z) &= \sum_{k=0}^{|M|-1} (D^{-1})^j_{k} \psi_{T^2}^{k,M}(z)\, .
\end{align}

The numbers of physical states are shown in Table~\ref{tab:Z3}, which are counted by the rank of $(M_\eta)_{k}^{j}$, as pointed out in Ref.~\cite{Abe:2014noa}.
\begin{table}[H]
    \centering
    \begin{tabular}{|c|c|c|c|c|c|c|} \hline
        \multicolumn{2}{|c|}{$M$} & 2 & 4 & 6 & 8 & 10  \\ \hline
        & 1 & 1 & 1 & 3 & 3 & 3\\
        $\eta$ & $\omega$ & 0 & 2 & 2 & 2 & 4\\
        & $\omega^2$ & 1 & 1 & 1 & 3 & 3\\ \hline
    \end{tabular}
    \caption{The numbers of $\mathbb{Z}_3$-untwisted and -twisted states on $T^2/\mathbb{Z}_3$, where $\eta$ denotes the $\mathbb{Z}_3$ eigenvalue. Only an even number of flux can be allowed in the vase of vanishing Wilson lines.}
    \label{tab:Z3}
\end{table}

Let us act the $\mathbb{Z}_3$-invariant operator on $|\psi\rangle_{\hat{T}^2}^{j,M}$:
\begin{align}
    \hat{U}_{\mathbb{Z}_3}^{(\lambda_1, \lambda_2)}|\psi\rangle_{\hat{T}^2}^{j,M}&=
    \left( e^{i\lambda_1 \hat{T}_1+i \lambda_2 \hat{T}_2}+ e^{i \lambda_1 \hat{T}_2 -i
\lambda_2(\hat{T}_1+\hat{T}_2)} + e^{-i\lambda_1 (\hat{T}_1+\hat{T}_2)  +i
\lambda_2 \hat{T}_1}
\right)|\psi\rangle_{T^2}^{j,M}
\nonumber\\
&= e^{-\pi i \frac{n_1 n_2}{M}}e^{-2\pi i\frac{n_2 j}{M}}|\psi\rangle_{T^2}^{j-n_1,M}
+ e^{\pi i\frac{(n_1 -n_2) n_2}{M}}
e^{-2\pi i\frac{(n_1 -n_2)j}{M}}|\psi\rangle_{T^2}^{j+n_2,M}
\nonumber\\
&+ e^{-\pi i\frac{(n_1 - n_2) n_1}{M}}
e^{2\pi i\frac{n_1 j}{M}}|\psi\rangle_{T^2}^{j+n_1 -n_2,M}\,,
\end{align}
 with $\lambda_i = n_i/M$, 
 where we use the Baker-Campbell-Hausdorff formula for $\hat{T}_1$ and $\hat{T}_2$ with $[\hat{T}_2, \hat{T}_1]=2\pi i M$ and 
 \begin{align}
     e^{i\frac{n}{M} \hat{T}_1}|\psi\rangle_{T^2}^{j,M}&=
     |\psi\rangle_{T^2}^{j-n,M}\,,
     \nonumber\\
     e^{i\frac{n}{M} \hat{T}_2}|\psi\rangle_{T^2}^{j,M}&=
     e^{-2\pi i\frac{n}{M}j}|\psi\rangle_{T^2}^{j-n,M}\,.
     \label{eq:T1T2trf}
 \end{align}

By utilizing Eqs.~\eqref{eq:Pstate} and~\eqref{eq:UPcom}, we arrive at the transformations of chiral zero modes under the non-invertible symmetry:
\begin{align}
    \hat{U}_{\mathbb{Z}_3}^{(\lambda_1, \lambda_2)}|\psi\rangle_{\mathbb{Z}_3, \eta}^{j,M}&=
e^{-\pi i \frac{n_1 n_2}{M}}e^{-2\pi i\frac{n_2 j}{M}}|\psi\rangle_{\mathbb{Z}_3, \eta}^{j-n_1,M}
+ e^{\pi i\frac{(n_1 -n_2) n_2}{M}}
e^{-2\pi i\frac{(n_1 -n_2)j}{M}}|\psi\rangle_{\mathbb{Z}_3, \eta}^{j+n_2,M}
\nonumber\\
&+ e^{-\pi i\frac{(n_1 - n_2) n_1}{M}}
e^{2\pi i\frac{n_1 j}{M}}|\psi\rangle_{\mathbb{Z}_3, \eta}^{j+n_1 -n_2,M}\,.
\label{eq:trfZ3}
\end{align}
When $n_1=n_2=1$, it is simplified as
\begin{align}
\hat{U}_{\mathbb{Z}_3}^{(1/M, 1/M)}|\psi\rangle_{\mathbb{Z}_3, \eta}^{j,M} &=
    \rho^j |\psi\rangle_{\mathbb{Z}_3, \eta}^{j,M} 
    +|\psi\rangle_{\mathbb{Z}_3, \eta}^{j+1,M}
    + \rho^{-\frac{1}{2}-j} |\psi\rangle_{\mathbb{Z}_3, \eta}^{j-1,M}\,,
\end{align}
with $\rho = e^{\frac{2\pi i}{M}}$.
Note that the state $|\psi\rangle_{\mathbb{Z}_3, \eta}^{j,M}$ is not an eigenstate under $\mathbb{Z}_3$. 
In the following analysis, we examine the non-invertible transformations of chiral zero modes on eigenstates of $\mathbb{Z}_3$.

For illustrative purposes, we discuss the transformations of chiral zero modes under the non-invertible symmetry in the $M=2$ case. 
When $M=2$, there is a $\mathbb{Z}_3$-untwisted state and a twisted state with $\eta=\omega^2$ shown in Table~\ref{tab:Z3}, which are written in the $\mathbb{Z}_3$ eigenstate basis by diagonalizing $ (M_\eta)_k^j$:
\begin{align}
 |\Psi\rangle_{\mathbb{Z}_3, \eta=1}^{M=2} &=\frac{e^{\frac{-\pi i}{4}}}{\sqrt{6}} |\psi\rangle_{T^2}^{0,2} 
 + \frac{3-\sqrt{3}}{6} |\psi\rangle_{T^2}^{1,2} 
    \,,
\nonumber\\
 |\Psi\rangle_{\mathbb{Z}_3, \eta=\omega^2}^{M=2} &= -\frac{e^{\frac{-\pi i}{4}}}{\sqrt{6}} |\psi\rangle_{T^2}^{0,2} 
 + \frac{3+\sqrt{3}}{6} |\psi\rangle_{T^2}^{1,2} 
    \,.
\end{align}
Under the non-invertible symmetry, they transform as
\small
\begin{align}
    \hat{U}_{\mathbb{Z}_3}^{(1/2, 1/2)}\,&:\,
    \begin{pmatrix}
   |\Psi\rangle_{\mathbb{Z}_3, \eta=1}^{M=2}\\ 
    |\Psi\rangle_{\mathbb{Z}_3, \eta=\omega^2}^{M=2}
    \end{pmatrix}
    \rightarrow 
    \frac{1}{\sqrt{3}}
    \begin{pmatrix}
      1 &  (1 - 3 i) - (1 - i) \sqrt{3} \\
      -(1 - 3 i) - (1 - i) \sqrt{3} & -1 
    \end{pmatrix}
    \begin{pmatrix}
   |\Psi\rangle_{\mathbb{Z}_3, \eta=1}^{M=2}\\ 
    |\Psi\rangle_{\mathbb{Z}_3, \eta=\omega^2}^{M=2}
    \end{pmatrix}
    \,,
\end{align}
\normalsize
for $n_1=n_2=1$ and we obtain a trivial representation for $n_1=n_2=2$, i.e., $\hat{U}_{\mathbb{Z}_3}^{(1, 1)}=3\,\mathbb{I}_2$. 
Hence, the non-invertible transformations mix the states with different $\mathbb{Z}_3$ eigenvalues. 
The representation matrix $\hat{U}_{\mathbb{Z}_3}^{(1/2, 1/2)}$ satisfies
\begin{align}
    \hat{U}_{\mathbb{Z}_3}^{(1/2, 1/2)}\hat{U}_{\mathbb{Z}_3}^{(1/2, 1/2)} = 3\,\mathbb{I}_2\,,
\end{align}
indicating that there exists the $\mathbb{Z}_2$ symmetry associated with the non-invertible transformation $\frac{1}{\sqrt{3}}\hat{U}_{\mathbb{Z}_3}^{(1/2, 1/2)}$.

Similarly, we examine the $\mathbb{Z}_3$-untwisted and -twisted states for $M=4,6$. 
It turns out that the non-invertible transformation mixes the states with different $\mathbb{Z}_3$ eigenvalues under the non-invertible transformations, as shown in Appendix \ref{app:Z3}.

\subsection{Non-invertible symmetries on $T^2/\mathbb{Z}_4$}
\label{sec:Z4}

In this section, we move to the $T^2/\mathbb{Z}_4$ orbifold. 
In contrast to the $T^2/\mathbb{Z}_3$ case, one can consider the odd number of flux quanta. 
On $T^2/\mathbb{Z}_4$ orbifold, chiral zero modes are obtained as
\begin{align}
    |\psi\rangle_{\mathbb{Z}_4,\eta}^{j,M} &= \sum_{k=0}^{|M|-1}(M_\eta)_{k}^{j} |\psi\rangle_{T^2}^{k,M}\,,
\end{align}
with 
\begin{align}
    (M_\eta)_{k}^{j} := \frac{1}{4}\left( \delta_k^j + \Bar{\eta}D^j_k + \Bar{\eta}^2 (\delta)^j_{-k} + \Bar{\eta}^3 (D^{-1})^j_k\right)\,,
\end{align}
where $\eta$ denotes the $\mathbb{Z}_4$ eigenvalues, i.e., $\eta=1,\omega, \omega^2, \omega^3$, and 
\begin{align}
    D_{k}^{j} &:= \frac{1}{\sqrt{M}}e^{2\pi i \frac{jk}{M}}\,,
\nonumber\\
    (D^{-1})_{k}^{j} &:= \frac{1}{\sqrt{M}}e^{- 2\pi i \frac{jk}{M}}\,. 
\end{align}
Here and in what follows, we focus on the case with $M>0$. 
Note that in the language of wave function, the $\mathbb{Z}_4$-twisted mode function $\psi_{T^2}^{j,M}(\omega^xz)$ with $x=1,2,3$ can be expanded by $\psi_{T^2}^{j,M}(z)$~\cite{Abe:2014noa}
\begin{align}
    \psi_{T^2}^{j,M}(\omega z) &= \sum_{k=0}^{M-1} D^j_{k} \psi_{T^2}^{k,M}(z)\,,
    \nonumber\\
    \psi_{T^2}^{j,M}(\omega^2 z) &= \sum_{k=0}^{M-1} \delta_{-j,k} \psi_{T^2}^{k,M}(z)\,,
    \nonumber\\
    \psi_{T^2}^{j,M}(\omega^3 z) &= \sum_{k=0}^{M-1} (D^{-1})^j_{k} \psi_{T^2}^{k,M}(z)\, .
\end{align}
The numbers of $\mathbb{Z}_4$-untwisted and twisted states are shown in Table~\ref{tab:Z3}, which are counted by the rank of $(M_\eta)_{k}^{j}$, as pointed out in Ref.~\cite{Abe:2014noa}.
\begin{table}[H]
    \centering
    \begin{tabular}{|c|c|c|c|c|c|c|c|c|c|c|c|} \hline
        \multicolumn{2}{|c|}{$M$} & 1 & 2 & 3 & 4 & 5 & 6 & 7 & 8 & 9 & 10  \\ \hline
        & 1 & 1 & 1 & 1 & 2 & 2 & 2 & 2 & 3 & 3 & 3\\
        $\eta$ & $\omega$ & 0 & 0 & 1 & 1 & 1 & 1 & 2 & 2 & 2 & 2\\
        & $\omega^2$ & 0 & 1 & 1 & 1 & 1 & 2 & 2 & 2 & 2 & 3\\ 
        & $\omega^3$ & 0 & 0 & 0 & 0 & 1 & 1 & 1 & 1 & 2 & 2\\ \hline
    \end{tabular}
    \caption{The numbers of $\mathbb{Z}_4$-untwisted and twisted states on $T^2/\mathbb{Z}_4$, where $\eta$ denotes the $\mathbb{Z}_4$ eigenvalue.}
    \label{tab:Z4}
\end{table}

Let us act the $\mathbb{Z}_4$-invariant operator on $|\psi\rangle_{\hat{T}^2}^{j,M}$:
\begin{align}
    \hat{U}_{\mathbb{Z}_4}^{(\lambda_1, \lambda_2)}|\psi\rangle_{T^2}^{j,M}&=
    \left( e^{i\lambda_1 \hat{T}_1 +i\lambda_2 \hat{T}_2 } + e^{-i\lambda_1 \hat{T}_2 + i\lambda_2 \hat{T}_1} + e^{-i\lambda_1 \hat{T}_1 - i\lambda_2 \hat{T}_2}
+ e^{i\lambda_1 \hat{T}_2 - i\lambda_2 \hat{T}_1}
\right)|\psi\rangle_{T^2}^{j,M}
\nonumber\\
&= e^{-\pi i \frac{n_1 n_2}{M}}e^{-2\pi i\frac{n_2 j}{M}}|\psi\rangle^{j-n_1,M}
+ e^{\pi i \frac{n_1 n_2}{M}}e^{2\pi i\frac{n_2 j}{M}}|\psi\rangle_{T^2}^{j-n_2,M}
\nonumber\\
&+ e^{-\pi i \frac{n_1 n_2}{M}}e^{2\pi i\frac{n_2 j}{M}}|\psi\rangle_{T^2}^{j+n_1,M}
+ e^{\pi i \frac{n_1 n_2}{M}}e^{-2\pi i\frac{n_2 j}{M}}|\psi\rangle_{T^2}^{j+n_2,M}\,,
\end{align}
 with $\lambda_i = n_i/M$, 
 where we use the Baker-Campbell-Hausdorff formula for $\hat{T}_1$ and $\hat{T}_2$ and Eq.~\eqref{eq:T1T2trf}.
 Then, together with Eqs.~\eqref{eq:Pstate} and~\eqref{eq:UPcom}, we arrive at the transformations of the chiral zero modes under the non-invertible symmetry:
\begin{align}
    \hat{U}_{\mathbb{Z}_4}^{(\lambda_1, \lambda_2)}|\psi\rangle_{\mathbb{Z}_4, \eta}^{j,M}&=
e^{-\pi i \frac{n_1 n_2}{M}}e^{-2\pi i\frac{n_2 j}{M}}|\psi\rangle_{\mathbb{Z}_4, \eta}^{j-n_1,M}
+ e^{\pi i \frac{n_1 n_2}{M}}e^{2\pi i\frac{n_2 j}{M}}|\psi\rangle_{\mathbb{Z}_4, \eta}^{j-n_2,M}
\nonumber\\
&+ e^{-\pi i \frac{n_1 n_2}{M}}e^{2\pi i\frac{n_2 j}{M}}|\psi\rangle_{\mathbb{Z}_4, \eta}^{j+n_1,M}
+ e^{\pi i \frac{n_1 n_2}{M}}e^{-2\pi i\frac{n_2 j}{M}}|\psi\rangle_{\mathbb{Z}_4, \eta}^{j+n_2,M}\,.
\label{eq:trfZ4}
\end{align}

For illustrative purposes, we discuss the transformations of chiral zero modes under the non-invertible symmetry in the $M=4$ case, where the numbers of $\mathbb{Z}_4$-untwisted and -twisted states are shown in Table~\ref{tab:Z4}. 
For other cases, we discuss the representation matrix of chiral zero modes under the non-invertible symmetry in Appendix \ref{app:Z4}. 
By diagonalizing $ (M_\eta)_k^j$, one can show zero-mode wave functions for each $\mathbb{Z}_4$ eigenvalue:
\begin{align}
    |\Psi\rangle_{\mathbb{Z}_4, \eta=1}^{1,M=4}&=\frac{1}{4}(|\psi\rangle_{T^2}^{0,4} + |\psi\rangle_{T^2}^{1,4}-|\psi\rangle_{T^2}^{2,4}+|\psi\rangle_{T^2}^{3,4})\,,
    \nonumber\\
    |\Psi\rangle_{\mathbb{Z}_4, \eta=1}^{2,M=4}&=\frac{1}{4}(|\psi\rangle_{T^2}^{0,4} - |\psi\rangle_{T^2}^{1,4}+3|\psi\rangle_{T^2}^{2,4}-|\psi\rangle_{T^2}^{3,4})\,,
    \nonumber\\
    |\Psi\rangle_{\mathbb{Z}_4, \eta=\omega}^{M=4}&=-\frac{1}{2}|\psi\rangle_{T^2}^{1,4} + \frac{1}{2}|\psi\rangle_{T^2}^{3,4}\,,  
    \nonumber\\
    |\Psi\rangle_{\mathbb{Z}_4, \eta=\omega^2}^{M=4}&=\frac{1}{4}(-|\psi\rangle_{T^2}^{0,4} + |\psi\rangle_{T^2}^{1,4}+|\psi\rangle_{T^2}^{2,4}+|\psi\rangle_{T^2}^{3,4})\,.
\end{align}
From Eq. \eqref{eq:trfZ4}, we find the following transformations
\begin{align}
    \begin{pmatrix}
   |\Psi\rangle_{\mathbb{Z}_4, \eta=1}^{1,M=4}\\ 
   |\Psi\rangle_{\mathbb{Z}_4, \eta=1}^{2,M=4}\\ 
   |\Psi\rangle_{\mathbb{Z}_4, \eta=\omega}^{M=4}\\ 
    |\Psi\rangle_{\mathbb{Z}_4, \eta=\omega^2}^{M=4}
    \end{pmatrix}
    \rightarrow 
    2\sqrt{2}
    \begin{pmatrix}
      0  & 0 & 0 & 1\\
      0  & 0 & 0 & -1\\
      0  & 0 & 0 & 0\\
      -1  & 0 & 0 & 0
    \end{pmatrix}
    \begin{pmatrix}
   |\Psi\rangle_{\mathbb{Z}_4, \eta=1}^{1,M=4}\\ 
   |\Psi\rangle_{\mathbb{Z}_4, \eta=1}^{2,M=4}\\ 
   |\Psi\rangle_{\mathbb{Z}_4, \eta=\omega}^{M=4}\\ 
    |\Psi\rangle_{\mathbb{Z}_4, \eta=\omega^2}^{M=4}
    \end{pmatrix}
    \,,
\end{align}
for $n_1=n_2=1,3$ and 
\begin{align}
    \begin{pmatrix}
   |\Psi\rangle_{\mathbb{Z}_4, \eta=1}^{1,M=4}\\ 
   |\Psi\rangle_{\mathbb{Z}_4, \eta=1}^{2,M=4}\\ 
   |\Psi\rangle_{\mathbb{Z}_4, \eta=\omega}^{M=4}\\ 
    |\Psi\rangle_{\mathbb{Z}_4, \eta=\omega^2}^{M=4}
    \end{pmatrix}
    \rightarrow 
    4
    \begin{pmatrix}
      1  & 0 & 0 & 0\\
      -2  & -1 & 0 & 0\\
      0  & 0 & -1 & 0\\
      0  & 0 & 0 & 1
    \end{pmatrix}
    \begin{pmatrix}
   |\Psi\rangle_{\mathbb{Z}_4, \eta=1}^{1,M=4}\\ 
   |\Psi\rangle_{\mathbb{Z}_4, \eta=1}^{2,M=4}\\ 
   |\Psi\rangle_{\mathbb{Z}_4, \eta=\omega}^{M=4}\\ 
    |\Psi\rangle_{\mathbb{Z}_4, \eta=\omega^2}^{M=4}
    \end{pmatrix}
    \,.
\end{align}
for $n_1=n_2=2$. 
In the latter $n_1=n_2=2$ case, there exists a $\mathbb{Z}_2$ symmetry associated with $\frac{1}{4}\hat{U}_{\mathbb{Z}_4}^{(1/2, 1/2)}$. 
In any case, the representation obeys the fusion rule \eqref{eq:fusionrulesZN}.

\subsection{Non-invertible symmetries on $T^2/\mathbb{Z}_6$}
\label{sec:Z6}

Finally, we discuss the $T^2/\mathbb{Z}_6$ orbifold. 
Under the $\mathbb{Z}_6$ action $z\rightarrow \omega z$ with $\omega=e^{\frac{2\pi i}{6}}$, 
the complex structure of $T^2$ is fixed at $\tau = \omega$.

Similarly to $T^2/\mathbb{Z}_3$ case, only the even number of flux quanta is allowed in the case of vanishing Wilson lines. 
On $T^2/\mathbb{Z}_6$ orbifold, chiral zero modes are obtained as
\begin{align}
    |\psi\rangle_{\mathbb{Z}_6,\eta}^{j,M} &= \sum_{k=0}^{|M|-1}(M_\eta)_{k}^{j} |\psi\rangle_{T^2}^{k,M}\,,
\end{align}
with 
\begin{align}
    (M_\eta)_{k}^{j} := \frac{1}{6}\sum_{x=0}^5 \Bar{\eta}^x (D^{(\omega^x)})^j_k\,,
\end{align}
where $\eta$ denotes the $\mathbb{Z}_6$ eigenvalues, i.e., $\eta=1,\omega, \omega^2,...,\omega^5$, and 
\begin{align}
    (D^{(\omega)})_{k}^{j} &:= \frac{1}{\sqrt{M}}e^{i\frac{\pi}{12}}e^{-i\frac{\pi}{M}k^2}e^{2\pi i\frac{jk}{M}}\,,
\nonumber\\
    (D^{(\omega^2)})_{k}^{j} &:= \frac{1}{\sqrt{M}}e^{-i\frac{\pi}{12}}e^{i\frac{\pi}{M}j^2}e^{2\pi i\frac{jk}{M}}\,,
\nonumber\\
    (D^{(\omega^3)})_{k}^{j} &:= (\delta)^j_{-k}\,,
\nonumber\\
    (D^{(\omega^4)})_{k}^{j} &:= \frac{1}{\sqrt{M}}e^{i\frac{\pi}{12}}e^{-i\frac{\pi}{M}k^2}e^{-2\pi i\frac{jk}{M}}\,,
\nonumber\\
    (D^{(\omega^5)})_{k}^{j} &:= \frac{1}{\sqrt{M}}e^{-i\frac{\pi}{12}}e^{i\frac{\pi}{M}j^2}e^{-2\pi i\frac{jk}{M}}\,.
\end{align}
Here and in what follows, we focus on the case with $M>0$. 
Note that in the language of wave function, the $\mathbb{Z}_6$-twisted mode functions $\psi_{T^2}^{j,M}(\omega^x z)$ with $x=1,2,3,4,5$ can be expanded by $\psi_{T^2}^{j,M}(z)$~\cite{Abe:2014noa}
\begin{align}
    \psi_{T^2}^{j,M}(\omega^x z) &= \sum_{k=0}^{M-1} (D^{(\omega^x)})_{k}^{j} \psi_{T^2}^{k,M}(z)\,.
\end{align}

The numbers of $\mathbb{Z}_6$-untwisted and twisted states are shown in Table~\ref{tab:Z3}, which are counted by the rank of $(M_\eta)_{k}^{j}$, as pointed out in Ref.~\cite{Abe:2014noa}.
\begin{table}[H]
    \centering
    \begin{tabular}{|c|c|c|c|c|c|c|} \hline
        \multicolumn{2}{|c|}{$M$} & 2 & 4 & 6 & 8 & 10  \\ \hline
        & 1 & 1 & 1 & 2 & 2 & 2\\
        & $\omega$ & 0 & 1 & 1 & 1 & 2\\
        & $\omega^2$ & 1 & 1 & 1 & 2 & 2\\ 
        $\eta$ & $\omega^3$ & 0 & 0 & 1 & 1 & 1\\ 
        & $\omega^4$ & 0 & 1 & 1 & 1 & 2\\ 
        & $\omega^5$ & 0 & 0 & 0 & 1 & 1\\ \hline
    \end{tabular}
    \caption{The numbers of $\mathbb{Z}_6$-untwisted and twisted states on $T^2/\mathbb{Z}_6$, where $\eta$ denotes the $\mathbb{Z}_6$ eigenvalue.}
    \label{tab:Z6}
\end{table}

Let us act the $\mathbb{Z}_6$-invariant operator on $|\psi\rangle_{\hat{T}^2}^{j,M}$:
\begin{align}
    \hat{U}_{\mathbb{Z}_6}^{(\lambda_1, \lambda_2)}|\psi\rangle_{T^2}^{j,M}=&
    \left( e^{i\lambda_1 \hat{T}_1 +i\lambda_2 \hat{T}_2} + e^{i\lambda_1 (\hat{T}_1-\hat{T}_2) + i\lambda_2 \hat{T}_1} + e^{-i\lambda_1 \hat{T}_2 + i\lambda_2 (\hat{T}_1-\hat{T}_2)}
+ e^{-i\lambda_1 \hat{T}_1 - i\lambda_2 \hat{T}_2}
\right.
\nonumber\\
&
\left.
+ e^{-i\lambda_1 (\hat{T}_1-\hat{T}_2) - i\lambda_2 \hat{T}_1}
+ e^{i\lambda_1 \hat{T}_2 - i\lambda_2 (\hat{T}_1 -\hat{T}_2)}
\right)|\psi\rangle_{T^2}^{j,M}
\nonumber\\
&= e^{-\pi i \frac{n_1 n_2}{M}}e^{-2\pi i\frac{n_2 j}{M}}|\psi\rangle_{T^2}^{j-n_1,M}
+ e^{\pi i \frac{(n_1  + n_2)n_1}{M}}e^{2\pi i\frac{n_1 j}{M}}|\psi\rangle_{T^2}^{j-n_1-n_2,M}
\nonumber\\
&+ e^{\pi i \frac{(n_1 + n_2)n_2}{M}}e^{2\pi i\frac{(n_1 + n_2) j}{M}}|\psi\rangle_{T^2}^{j-n_2,M}
+ e^{-\pi i \frac{n_1n_2}{M}}e^{2\pi i\frac{n_2 j}{M}}|\psi\rangle_{T^2}^{j+n_1,M}
\nonumber\\
&+ e^{\pi i \frac{(n_1 + n_2)n_1}{M}}e^{-2\pi i\frac{n_1 j}{M}}|\psi\rangle_{T^2}^{j+n_1+n_2,M}
+ e^{\pi i \frac{(n_1 + n_2)n_2}{M}}e^{-2\pi i\frac{(n_1 +n_2) j}{M}}|\psi\rangle_{T^2}^{j+n_2,M}\,,
\end{align}
 with $\lambda_i = n_i/M$, 
 where we use the Baker-Campbell-Hausdorff formula for $\hat{T}_1$ and $\hat{T}_2$ and Eq.~\eqref{eq:T1T2trf}.
 Then, together with Eqs.~\eqref{eq:Pstate} and~\eqref{eq:UPcom}, we arrive at the transformations of the chiral zero modes under the non-invertible symmetry:
\begin{align}
    \hat{U}_{\mathbb{Z}_6}^{(\lambda_1, \lambda_2)}|\psi\rangle_{\mathbb{Z}_6, \eta}^{j,M}&=
e^{-\pi i \frac{n_1 n_2}{M}}e^{-2\pi i\frac{n_2 j}{M}}|\psi\rangle_{\mathbb{Z}_6, \eta}^{j-n_1,M}
+ e^{\pi i \frac{(n_1  + n_2)n_1}{M}}e^{2\pi i\frac{n_1 j}{M}}|\psi\rangle_{\mathbb{Z}_6, \eta}^{j-n_1-n_2,M}
\nonumber\\
&+ e^{\pi i \frac{(n_1 + n_2)n_2}{M}}e^{2\pi i\frac{(n_1 + n_2) j}{M}}|\psi\rangle_{\mathbb{Z}_6, \eta}^{j-n_2,M}
+ e^{-\pi i \frac{n_1n_2}{M}}e^{2\pi i\frac{n_2 j}{M}}|\psi\rangle_{\mathbb{Z}_6, \eta}^{j+n_1,M}
\nonumber\\
&+ e^{\pi i \frac{(n_1 + n_2)n_1}{M}}e^{-2\pi i\frac{n_1 j}{M}}|\psi\rangle_{\mathbb{Z}_6, \eta}^{j+n_1+n_2,M}
+ e^{\pi i \frac{(n_1 + n_2)n_2}{M}}e^{-2\pi i\frac{(n_1 +n_2) j}{M}}|\psi\rangle_{\mathbb{Z}_6, \eta}^{j+n_2,M}\,.
\end{align}

For illustrative purposes, we discuss the transformations of chiral zero modes under the non-invertible symmetry in the $M=2$ case, where the numbers of $\mathbb{Z}_6$-untwisted and -twisted states are shown in Table~\ref{tab:Z6}. 
For other $M$ cases, the non-invertible transformations can be analyzed in Appendix \ref{app:Z6}. 
By diagonalizing $ (M_\eta)_k^j$, one can show zero-mode wave functions for each $\mathbb{Z}_6$ eigenvalue:
\begin{align}
 |\Psi\rangle^{1,2}_{\mathbb{Z}_6,\eta=1} &=\frac{(1-i)(1+\sqrt{3})}{2}|\psi\rangle_{T^2}^{0,2}
 + |\psi\rangle_{T^2}^{1,2}
    \,,
\nonumber\\
|\Psi\rangle^{1,2}_{\mathbb{Z}_6,\eta=\omega^2} &=-\frac{1+i}{2\sqrt{3}}|\psi\rangle_{T^2}^{0,2}
 + \frac{1+\sqrt{3}}{2\sqrt{3}}|\psi\rangle_{T^2}^{1,2}     \,.
\end{align}

We find that their non-invertible transformations with $n_1 = n_2=1$ are given by
\begin{align}
\hat{U}_{\mathbb{Z}_6}^{(1/2, 1/2)}\,:\, 
    \begin{pmatrix}
        |\Psi\rangle^{1,2}_{\mathbb{Z}_6,\eta=1}\\
        |\Psi\rangle^{1,2}_{\mathbb{Z}_6,\eta=\omega^2}\\
    \end{pmatrix}
    &\rightarrow 2\sqrt{3}
    \begin{pmatrix}
        -1 & 0 \\
        -\frac{1-i}{3} & 1\\
    \end{pmatrix}
    \begin{pmatrix}
        |\Psi\rangle^{1,2}_{\mathbb{Z}_6,\eta=1}\\
        |\Psi\rangle^{1,2}_{\mathbb{Z}_6,\eta=\omega^2}\\
    \end{pmatrix}
    .
\end{align}
In this specific example, $\mathbb{Z}_6$ eigenstates mix with each other under the non-invertible transformations. 
Furthermore, there exists a $\mathbb{Z}_2$ symmetry associated with $\widetilde{\hat{U}}_{\mathbb{Z}_6}^{(1/2, 1/2)}=\frac{1}{2\sqrt{3}}\hat{U}_{\mathbb{Z}_6}^{(1/2, 1/2)}$.


\section{Conclusions}
\label{sec:con}

We have studied the non-invertible symmetries in orbifold compactifications with magnetic flux background. 
In particular, we have revealed fusion rules of discrete isometry operators, which are invariant under the $\mathbb{Z}_N$ orbifold twist, and their representations explicitly on zero-mode wave functions. 
These zero modes may correspond to generations of quarks and leptons 
in 4D low-energy effective field theory.
Hence, the non-invertible symmetries describe flavor symmetries among quarks and leptons. We mainly focused on the chiral zero modes, but our methods can be applied to Kaluza-Klein states as well.

Various representations of our fusion rules have been found including trivial ones. 
One of representations on $T^2/\mathbb{Z}_2$ with even flux corresponds to 
representations of $D_4$ symmetry. 
On the other hand, there is no invertible symmetry in the case of odd flux, but the non-invertible symmetry exists. 
We find that the non-invertible symmetry controls the coupling selection rules of zero modes in specific examples, but it is also important to classify representations of our fusion rules. 
Moreover, it is interesting to apply those representations to particle physics as flavor physics, although the $D_4$ flavor symmetry was used in the bottom-up approach of many flavor models \cite{Altarelli:2010gt,Ishimori:2010au,Kobayashi:2022moq,Hernandez:2012ra,King:2013eh}. 
However, it was known that in general, a non-invertible $0$-form symmetry in $d$ dimensions can appear only if there is also a $d-2$ form symmetry. 
We have found the non-invertible $0$-form symmetry acting on zero modes in orbifold compactifications with background magnetic fluxes, but it is important to understand the structure of low-energy effective action. 
We will leave them for future work.

We have discussed magnetized D-brane models.
The intersecting D-brane models are T-dual to magnetized D-brane models.
Indeed, they have similar coupling selection rules \cite{Cvetic:2003ch,Abel:2003vv,Abel:2003yx,Higaki:2005ie} and flavor symmetries \cite{Marchesano:2013ega}.
The open string between intersecting D-branes has a twisted boundary condition.
Its ground state is generated by the twist field from the untwisted ground state 
like heterotic string theory on orbifolds \cite{Hamidi:1986vh,Dixon:1986qv}.
The non-invertible symmetry of twist fields on $S^1/\mathbb{Z}_2$ was studied in Refs.~\cite{Heckman:2024obe,Kaidi:2024wio}.
The fusion rule and transformations of operators are the same 
as our results on $T^2/\mathbb{Z}_2$, i.e., the fusion rule of $\{\hat{U}_{\hat{Y}}, \hat{U}_{\hat{P}}\}$ and transformation of wave functions on 
$T^2/\mathbb{Z}_2$.
The result in Ref.~\cite{Kaidi:2024wio} shows that  
the $D_4$ symmetry appears.
Such $D_4$ symmetry has already been derived in 
conformal field theory on $S^1/\mathbb{Z}_2$ \cite{Dijkgraaf:1987vp} and 
matter flavor symmetry in heterotic orbifold models \cite{Kobayashi:2004ya,Kobayashi:2006wq}.  
Hence, it is of interest to study non-invertible symmetries 
of twist fields in intersecting D-brane models and 
heterotic string theory on $T^6/\mathbb{Z}_N$ in order to understand 
non-invertible flavor symmetries.
We would study them elsewhere.

Finally, we comment on the breaking of non-invertible symmetry in toroidal orbifolds. 
When we consider resolutions of toroidal orbifolds, orbifold groups are in general broken down, 
depending on the vacuum expectation values of blow-up modes which control the size of blow-up radius. 
Hence, the dynamics of blow-up modes will determine whether the non-invertible symmetries remain in the low-energy effective action. 
We will leave a comprehensive study about breaking of non-invertible symmetries for future work.

\acknowledgments

We thank J. Kaidi and R. Yokokura for helpful comments. 
The authors thank the Yukawa Institute for Theoretical Physics 
at Kyoto University, where this work was initiated 
during the YITP-W-24-09 on "Progress in Particle Physics 2024". 
This work was supported in part JSPS KAKENHI Grant Numbers JP23H04512 (H.O) and JP23K03375 (T.K.).

\appendix

\section{Transformations of chiral zero modes under non-invertible symmetries on $T^2/\mathbb{Z}_N$}
\label{app}

In this appendix, we summarize the transformations of chiral zero modes under non-invertible symmetries on $T^2/\mathbb{Z}_N$ with $N=3,4,6$. 

\subsection{$T^2/\mathbb{Z}_3$}
\label{app:Z3}

For $M=2$, the chiral zero modes and their transformations under the non-invertible symmetry are presented in Sec. \ref{sec:Z3}. 
Hence, we show $M=4,6$ cases in the following. 

\paragraph{$M=4$}\,\\

When $M=4$, the numbers of $\mathbb{Z}_3$-untwisted and -twisted states are shown in Table~\ref{tab:Z3}. 
By diagonalizing $ (M_\eta)_k^j$, 
zero-mode wave functions for each $\mathbb{Z}_3$ eigenvalue are given by
\begin{align}
 |\Psi\rangle_{\mathbb{Z}_3, \eta=1}^{M=4} &=\frac{1}{36}\left( -1 + e^{\frac{3\pi i}{4}} -5 e^{\frac{5\pi i}{6}} -5 e^{\frac{11\pi i}{12}}
 - \left( e^{\frac{5\pi i}{6}} + e^{\frac{11\pi i}{12}}\right) \right) |\psi\rangle_{T^2}^{0,4} 
 \nonumber\\
 &\,\,\,\,+ \frac{1}{6} \left( |\psi\rangle_{T^2}^{1,4} + |\psi\rangle_{T^2}^{3,4}\right)
+ \frac{2+i - \sqrt{3}}{12(1+ e^{\frac{\pi i}{12}})} |\psi\rangle_{T^2}^{2,4}
    \,,
    \nonumber\\
|\Psi\rangle_{\mathbb{Z}_3, \eta=\omega}^{1,M=4} &= 
 -\frac{1+\sqrt{2}-i}{6(2+\sqrt{2})} |\psi\rangle_{T^2}^{0,4} 
 -\frac{1}{3} |\psi\rangle_{T^2}^{1,4} 
 -\left( \frac{i}{3} +\frac{1+\sqrt{2}-i}{6(2+\sqrt{2})}\right) |\psi\rangle_{T^2}^{2,4} 
 +\frac{2}{3} |\psi\rangle_{T^2}^{3,4} 
    \,,
    \nonumber\\
|\Psi\rangle_{\mathbb{Z}_3, \eta=\omega}^{2,M=4} &= 
 -\frac{i}{3\sqrt{2}} |\psi\rangle_{T^2}^{0,4} 
 +\frac{i + e^{\frac{3\pi i}{4}}}{6} |\psi\rangle_{T^2}^{1,4} 
 +\frac{2 + \sqrt{2}}{6} |\psi\rangle_{T^2}^{2,4} 
 +\frac{i + e^{\frac{3\pi i}{4}}}{6} |\psi\rangle_{T^2}^{3,4} 
    \,,
\nonumber\\
|\Psi\rangle_{\mathbb{Z}_3, \eta=\omega^2}^{M=4} &= \frac{(i+\sqrt{3}) (-1 + e^{\frac{5\pi i}{12}})}{12} |\psi\rangle_{T^2}^{0,4} 
+  \frac{1}{6} \left( |\psi\rangle_{T^2}^{1,4} + |\psi\rangle_{T^2}^{3,4}\right)
+ \frac{(i+\sqrt{3}) (1 + e^{\frac{5\pi i}{12}})}{12}  |\psi\rangle_{T^2}^{2,4}
    \,.
\end{align}

From Eq. \eqref{eq:trfZ3}, we find the following representaiton matrix of chiral matters:
\small
\begin{align}
    \begin{pmatrix}
1.288 - 0.644i & -0.122 - 0.455i & 0.283 - 0.051i & 0.173 - 0.644i \\
1.429 + 0.130i & -0.333 - 0.667i & -0.098 + 0.983i & 0.514 - 1.602i \\
0.358 + 0.274i & 1.374 - 0.098i & -0.609 + 1.138i & -0.220 + 1.668i \\
-0.644 + 0.173i & 0.455 + 0.122i & 0.522 + 0.189i & -0.345 + 0.173i
    \end{pmatrix}
    \,,
\end{align}
\normalsize
for $n_1=n_2=1$, 
\begin{align}
    \begin{pmatrix}
1 & 0 & 0 & 0 \\
0 & -3 & -2 \left( (-1 + i) + \sqrt{2} \right) & 0 \\
0 & 0 & 1 & 0 \\
0 & 0 & 0 & 1 \\
    \end{pmatrix}
    \,,
\end{align}
for $n_1=n_2=2$, and 
\small
\begin{align}
    \begin{pmatrix}
1.288 - 0.644i & 0.122 + 0.455i & -0.122 + 0.260i & 0.173 - 0.644i \\
-1.303 - 0.602i & 0.333 + 0.667i & 0.960 - 0.236i & 1.246 + 1.130i \\
0.358 + 0.274i & -1.374 + 0.098i & -1.276 - 0.195i & -0.220 + 1.668i \\
-0.644 + 0.173i & -0.455 - 0.122i & 0.455 - 0.317i & -0.345 + 0.173i  
\end{pmatrix}
    \,,
\end{align}
\normalsize
for $n_1=n_2=3$. 
Here, we evaluate the transformation in the basis of
\begin{align}
\{|\Psi\rangle_{\mathbb{Z}_3, \eta=1}^{M=4}, |\Psi\rangle_{\mathbb{Z}_3, \eta=\omega}^{1,M=4},|\Psi\rangle_{\mathbb{Z}_3, \eta=\omega}^{2,M=4},|\Psi\rangle_{\mathbb{Z}_3, \eta=\omega^2}^{M=4}\}
\,,
\end{align} 
and numerically show the value of representation matrix for $n_1=n_2=1,3$. 
One can obtain the similar transformations for other $\{n_1,n_2\}$, 
and the fusion rule \eqref{eq:fusionrulesZN} is satisfied. 

\paragraph{$M=6$}\,\\

When $M=6$, the numbers of $\mathbb{Z}_3$-untwisted and -twisted states are shown in Table~\ref{tab:Z3}. 
By diagonalizing $ (M_\eta)_k^j$, 
zero-mode wave functions for each $\mathbb{Z}_3$ eigenvalue are numerically estimated as
\begin{align}
 |\Psi\rangle_{\mathbb{Z}_3, \eta=1}^{M=6} 
 &\simeq(0.0246 + 0.298i) |\psi\rangle_{T^2}^{0,6} 
 +(0.0284 + 0.553i)|\psi\rangle_{T^2}^{1,6} + (0.345 - 0.457 i)|\psi\rangle_{T^2}^{2,6}
\nonumber\\
&+(-0.156 - 0.0111 i)|\psi\rangle_{T^2}^{3,6}
-(0.372 + 0.337 i)|\psi\rangle_{T^2}^{4,6}
    \,,
    \nonumber\\
|\Psi\rangle_{\mathbb{Z}_3, \eta=1}^{2,M=6} 
&\simeq 
(0.414 - 0.111 i) |\psi\rangle_{T^2}^{0,6} 
 +0.303|\psi\rangle_{T^2}^{1,6} + 0.151(1+i)|\psi\rangle_{T^2}^{2,6}
\nonumber\\
&+(-0.151 + 0.262 i)|\psi\rangle_{T^2}^{3,6}
-0.303(1+i)|\psi\rangle_{T^2}^{4,6}+0.635|\psi\rangle_{T^2}^{5,6}
    \,,
    \nonumber\\
|\Psi\rangle_{\mathbb{Z}_3, \eta=1}^{3,M=6} 
&\simeq 
(0.550 - 0.144 i) |\psi\rangle_{T^2}^{0,6} 
 +(-0.0332 + 0.0664 i)|\psi\rangle_{T^2}^{1,6} + (0.392 + 0.263 i)|\psi\rangle_{T^2}^{2,6}
\nonumber\\
&+(0.227 - 0.486 i)|\psi\rangle_{T^2}^{3,6}
+(0.256 + 0.308i)|\psi\rangle_{T^2}^{4,6}
   \,,
    \nonumber\\
|\Psi\rangle_{\mathbb{Z}_3, \eta=\omega}^{1,M=6} 
&\simeq 
(0.250 - 0.208 i) |\psi\rangle_{T^2}^{0,6} 
 +(-0.570 + 0.263i)|\psi\rangle_{T^2}^{1,6} 
 \nonumber\\
&+(-0.0570 + 0.625 i)|\psi\rangle_{T^2}^{3,6}
+(0.305 + 0.112 i)|\psi\rangle_{T^2}^{4,6}
   \,,
    \nonumber\\
|\Psi\rangle_{\mathbb{Z}_3, \eta=\omega}^{2,M=6} 
&\simeq 
(0.181 - 0.0486 i) |\psi\rangle_{T^2}^{0,6} 
 +0.363|\psi\rangle_{T^2}^{1,6} +0.265(1+i)|\psi\rangle_{T^2}^{2,6}
 \nonumber\\
&+(-0.181 + 0.314 i)|\psi\rangle_{T^2}^{3,6}
-0.133(1+i)|\psi\rangle_{T^2}^{4,6}-0.725|\psi\rangle_{T^2}^{5,6}
   \,,
    \nonumber\\
|\Psi\rangle_{\mathbb{Z}_3, \eta=\omega^2}^{M=6} 
&\simeq 
(0.495 - 0.133 i) |\psi\rangle_{T^2}^{0,6} 
 -0.265|\psi\rangle_{T^2}^{1,6} -0.363(1+i)|\psi\rangle_{T^2}^{2,6}
 \nonumber\\
&+(0.133 - 0.230 i)|\psi\rangle_{T^2}^{3,6}
-0.363(1+i)|\psi\rangle_{T^2}^{4,6}-0.265|\psi\rangle_{T^2}^{5,6}
   \,.
\end{align}

From Eq. \eqref{eq:trfZ3}, we find the following representation matrix of chiral matters:
\small
 \begin{align}
 \begin{pmatrix}
-0.374 + 1.724 i & -0.264 + 0.414 i & -0.609 + 0.341 i & 0.099 - 0.047 i & -0.131 + 0.672 i & -0.226 - 0.010 i \\
0.278 + 0.404 i & 0.374 - 0.606 i & 1.558 - 0.473 i & -0.383 + 0.034 i & -0.430 + 0.238 i & -0.336 + 0.047 i \\
-0.274 - 0.053 i & 1.238 + 0.789 i & 0.167 - 0.951 i & 0.311 - 0.589 i & 0.038 + 0.181 i & 0.051 + 0.469 i \\
-0.172 + 0.052 i & 0.562 + 0.082 i & 0.355 + 0.234 i & 0.114 - 0.636 i & -0.338 - 0.814 i & -0.205 + 0.189 i \\
0.352 + 0.638 i & 0.036 - 0.170 i & -0.312 - 0.263 i & -0.405 + 0.695 i & -0.342 + 0.697 i & -0.885 - 0.168 i \\
-0.113 + 0.325 i & 0.351 + 0.486 i & 0.172 - 0.399 i & -0.311 - 0.417 i & 0.491 + 0.172 i & 0.061 - 0.228 i \\
    \end{pmatrix}
    \,,
 \end{align}
\normalsize
for $n_1=n_2=1$,
\small
 \begin{align}
 \begin{pmatrix}
0.234 - 0.508 i & -0.347 - 0.764 i & -0.273 + 0.374 i & -0.707 + 0.059 i & -0.553 + 0.444 i & 0.172 - 0.856 i \\
-0.511 + 0.234 i & -0.234 + 0.810 i & 0.454 - 0.107 i & -0.133 + 0.053 i & -0.080 + 0.118 i & 1.141 - 0.643 i \\
0.072 + 0.357 i & 0.476 + 0.269 i & 0 - 0.302 i & -0.856 + 0.494 i & -0.433 + 0.888 i & 0 + 0.723 i \\
-0.303 - 0.641 i & -0.020 - 0.142 i & 0 - 0.988 i & 0 - 0.866 i & 0.484 + 0.718 i & 0 \\
-0.108 + 0.701 i & -0.062 + 0.129 i & -0.552 + 0.819 i & 0.484 - 0.718 i & 0 + 0.866 i & 0 \\
-0.449 + 1.015 i & -0.321 - 1.068 i & 0.627 - 0.362 i & 0 & 0 & 0 \\
    \end{pmatrix}
    \,,
 \end{align}
\normalsize
for $n_1=n_2=2$, 
\small
 \begin{align}
 \begin{pmatrix}
0.190 & -0.075 - 0.023 i & -0.488 - 0.230 i & 1.061 - 0.736 i & 0.319 - 0.049 i & 0.078 + 0.942 i \\
-0.075 + 0.023 i & 0.562 & -0.108 - 0.015 i & 0.161 - 0.205 i & -1.439 + 0.701 i & 0.071 + 0.177 i \\
-0.488 + 0.230 i & -0.108 + 0.015 i & -0.174 & -0.482 + 0.835 i & 0 & 0.659 + 1.141 i \\
1.061 + 0.736 i & 0.161 + 0.205 i & -0.482 - 0.835 i & -0.577 & 0 & 0 \\
0.319 + 0.049 i & -1.439 - 0.701 i & 0 & 0 & -0.577 & 0 \\
0.078 - 0.942 i & 0.071 - 0.177 i & 0.659 - 1.141 i & 0 & 0 & 0.577 \\
    \end{pmatrix}
    \,,
 \end{align}
\normalsize
for $n_1=n_2=3$, 
\small
 \begin{align}
 \begin{pmatrix}
-0.323 + 0.457 i & -0.052 - 0.560 i & 0.346 - 0.117 i & 0.404 + 0.583 i & -0.661 - 0.257 i & -1.104 - 0.119 i \\
-0.835 + 0.081 i & 0.584 - 0.608 i & 0.471 + 0.278 i & 0.113 + 0.088 i & -0.143 - 0.010 i & 0.765 + 0.812 i \\
0.188 - 0.424 i & 0.134 + 0.446 i & -0.262 + 0.151 i & 0.856 + 0.494 i & -0.986 + 0.069 i & 0.627 - 0.362 i \\
-0.404 + 0.583 i & -0.113 + 0.088 i & -0.856 + 0.494 i & -0.750 + 0.433 i & -0.380 + 0.778 i & 0 \\
-0.661 + 0.257 i & -0.143 + 0.010 i & -0.986 - 0.069 i & 0.864 + 0.060 i & 0.750 - 0.433 i & 0 \\
0.827 + 0.279 i & 1.127 - 0.666 i & -0.627 - 0.362 i & 0 & 0 & 0 \\
    \end{pmatrix}
    \,,
 \end{align}
\normalsize
for $n_1=n_2=4$, 
\small
 \begin{align}
 \begin{pmatrix}
-1.459 - 1.137 i & -0.009 - 0.446 i & -0.235 + 0.181 i & 0.059 - 0.132 i & 0.321 + 0.512 i & 0.285 + 0.057 i \\
-0.521 - 0.478 i & 0.560 + 0.358 i & 0.522 - 1.576 i & -0.253 + 0.299 i & 0.121 - 0.336 i & 0.214 + 0.380 i \\
-0.330 + 0.207 i & 1.453 - 0.763 i & 0.899 + 0.779 i & 0.285 + 0.494 i & 0.079 + 0.039 i & -0.241 \\
0.071 + 0.043 i & 0.167 - 0.374 i & -0.571 & 0.577 + 0.500 i & -0.758 - 0.092 i & -0.289 - 0.500 i \\
0.502 + 0.148 i & 0.433 - 0.175 i & 0.073 + 0.049 i & 0.199 + 0.737 i & -0.577 - 0.500 i & -0.040 + 0.576 i \\
-0.338 - 0.150 i & 0.356 - 0.107 i & 0.121 + 0.209 i & 0.577 & 0.479 - 0.323 i & 0 \\
    \end{pmatrix}
    \,,
 \end{align}
\normalsize
for $n_1=n_2=5$. 

Here, we evaluate the transformations in the basis of
\begin{align}
\{|\Psi\rangle_{\mathbb{Z}_3, \eta=1}^{1,M=6}, |\Psi\rangle_{\mathbb{Z}_3, \eta=1}^{2,M=6},|\Psi\rangle_{\mathbb{Z}_3, \eta=1}^{3,M=6},|\Psi\rangle_{\mathbb{Z}_3, \eta=\omega}^{1,M=6},|\Psi\rangle_{\mathbb{Z}_3, \eta=\omega}^{2,M=6},|\Psi\rangle_{\mathbb{Z}_3, \eta=\omega^2}^{M=6}\}
\,,
\end{align}
and we numerically show the transformation matrix. 
Furthermore, there exists a $\mathbb{Z}_2$ symmetry associated with $\widetilde{\hat{U}}_{\mathbb{Z}_3}^{(1/2, 1/2)}=\frac{1}{\sqrt{3}}\hat{U}_{\mathbb{Z}_3}^{(1/2, 1/2)}$. 
One can obtain the similar transformations for other $\{n_1,n_2\}$, and the fusion rule \eqref{eq:fusionrulesZN} is satisfied.

\subsection{$T^2/\mathbb{Z}_4$}
\label{app:Z4}

In the following, we show the transformations of chiral zero modes under the non-invertible symmetry up to $M=6$.

\paragraph{$M=1$}\,\\

When $M=1$, there is only a $\mathbb{Z}_4$-untwisted state:
\begin{align}
    |\Psi\rangle_{\mathbb{Z}_4, \eta=1}^{M=1}=|\psi\rangle_{T^2}^{0,1}\,.
\end{align}
From Eq. \eqref{eq:trfZ4}, the non-invertible transformation can be discussed for $n_1=n_2=0$, which does not cause the transformation of the state.

\paragraph{$M=2$}\,\\

When $M=2$, there is a $\mathbb{Z}_4$-untwisted state and $\mathbb{Z}_4$-twisted state with the eigenvalue $\eta=\omega^2$:
\begin{align}
    |\Psi\rangle_{\mathbb{Z}_4, \eta=1}^{M=2}&=\frac{1}{2\sqrt{2}}|\psi\rangle_{T^2}^{0,2} + \frac{2-\sqrt{2}}{4}|\psi\rangle_{T^2}^{1,2}\,,
    \nonumber\\
    |\Psi\rangle_{\mathbb{Z}_4, \eta=\omega^2}^{M=2}&=-\frac{1}{2\sqrt{2}}|\psi\rangle_{T^2}^{0,2} + \frac{2+\sqrt{2}}{4}|\psi\rangle_{T^2}^{1,2}\,.    
\end{align}
From Eq. \eqref{eq:trfZ4}, we find that the non-invertible transformation is described by a trivial representation $\hat{U}_{\mathbb{Z}_4}^{(\lambda_1, \lambda_2)}=0$ with $n_1=n_2=1$. 
On the other hand, when $n_1=1$ and $n_2=0$, 
one can obtain the non-invertible transformation:
\begin{align}
    \begin{pmatrix}
   |\Psi\rangle_{\mathbb{Z}_4, \eta=1}^{M=2}\\ 
    |\Psi\rangle_{\mathbb{Z}_4, \eta=\omega^2}^{M=2}
    \end{pmatrix}
    \rightarrow 
    \sqrt{2}
    \begin{pmatrix}
      1 +\sqrt{2} &~~  -1+\sqrt{2}\\
      1+\sqrt{2}  &~~ -1+\sqrt{2} 
    \end{pmatrix}
    \begin{pmatrix}
   |\Psi\rangle_{\mathbb{Z}_4, \eta=1}^{M=2}\\ 
    |\Psi\rangle_{\mathbb{Z}_4, \eta=\omega^2}^{M=2}
    \end{pmatrix}
    \,,
\end{align}
whose representation obeys the fusion rule \eqref{eq:fusionrulesZN}.

\paragraph{$M=3$}\,\\

When $M=3$, the numbers of $\mathbb{Z}_4$-untwisted and -twisted states are shown in Table~\ref{tab:Z4}. 
By diagonalizing $ (M_\eta)_k^j$, 
zero-mode wave functions for each $\mathbb{Z}_4$ eigenvalue are given by
\begin{align}
    |\Psi\rangle_{\mathbb{Z}_4, \eta=1}^{M=3}&=\frac{1}{2\sqrt{3}}|\psi\rangle_{T^2}^{0,3} + \frac{3-\sqrt{3}}{12}|\psi\rangle_{T^2}^{1,3} + \frac{3-\sqrt{3}}{12}|\psi\rangle_{T^2}^{2,3}\,,
    \nonumber\\
    |\Psi\rangle_{\mathbb{Z}_4, \eta=\omega}^{M=3}&=-\frac{1}{2}|\psi\rangle_{T^2}^{1,3} + \frac{1}{2}|\psi\rangle_{T^2}^{2,3}\,,  
    \nonumber\\
    |\Psi\rangle_{\mathbb{Z}_4, \eta=\omega^2}^{M=3}&=-\frac{1}{2\sqrt{3}}|\psi\rangle_{T^2}^{0,3} + \frac{3+\sqrt{3}}{12}|\psi\rangle_{T^2}^{1,3} + \frac{3+\sqrt{3}}{12}|\psi\rangle_{T^2}^{2,3}\,.
\end{align}
From Eq. \eqref{eq:trfZ4}, we find the-invertible transformation for $n_1=n_2=1$:
\begin{align}
    \begin{pmatrix}
   |\Psi\rangle_{\mathbb{Z}_4, \eta=1}^{M=3}\\ 
   |\Psi\rangle_{\mathbb{Z}_4, \eta=\omega}^{M=3}\\ 
    |\Psi\rangle_{\mathbb{Z}_4, \eta=\omega^2}^{M=3}
    \end{pmatrix}
    \rightarrow 
    \frac{1}{2}
    \begin{pmatrix}
      1 -\sqrt{3} & 0 &  3(-1+\sqrt{3})\\
      0 & -2 & 0\\
      -3(1+\sqrt{3}) & 0 & 1+\sqrt{3} 
    \end{pmatrix}
    \begin{pmatrix}
   |\Psi\rangle_{\mathbb{Z}_4, \eta=1}^{M=3}\\ 
   |\Psi\rangle_{\mathbb{Z}_4, \eta=\omega}^{M=3}\\ 
    |\Psi\rangle_{\mathbb{Z}_4, \eta=\omega^2}^{M=3}
    \end{pmatrix}
    \,.
\end{align}
In the same way, when $n_1=1$ and $n_2=2$, one can obtain the invertible transformation:
\begin{align}
    \begin{pmatrix}
   |\Psi\rangle_{\mathbb{Z}_4, \eta=1}^{M=3}\\ 
   |\Psi\rangle_{\mathbb{Z}_4, \eta=\omega}^{M=3}\\ 
    |\Psi\rangle_{\mathbb{Z}_4, \eta=\omega^2}^{M=3}
    \end{pmatrix}
    \rightarrow 
    \frac{1}{4}
    \begin{pmatrix}
      (1+i)(-2+\sqrt{3}-i) & 0 &  -3(1-i)+\sqrt{3}(3-i)\\
      0 & 4(1- e^{\pi/3}) & 0\\
      -3(1+i)-\sqrt{3}(3+i) & 0 & (-1+i)(2+\sqrt{3}-i) 
    \end{pmatrix}
    \begin{pmatrix}
   |\Psi\rangle_{\mathbb{Z}_4, \eta=1}^{M=3}\\ 
   |\Psi\rangle_{\mathbb{Z}_4, \eta=\omega}^{M=3}\\ 
    |\Psi\rangle_{\mathbb{Z}_4, \eta=\omega^2}^{M=3}
    \end{pmatrix}
    \,.
\end{align}
The representation obeys the fusion rule \eqref{eq:fusionrulesZN}.

\paragraph{$M=4$}\,\\

In this case, we discuss the transformation of chiral zero modes under the non-invertible symmetry in Sec. \ref{sec:Z4}.

\paragraph{$M=5$}\,\\

When $M=5$, the numbers of $\mathbb{Z}_4$-untwisted and -twisted states are shown in Table~\ref{tab:Z4}. 
By diagonalizing $ (M_\eta)_k^j$, 
zero-mode wave functions for each $\mathbb{Z}_4$ eigenvalue are given by
\begin{align}
    |\Psi\rangle_{\mathbb{Z}_4, \eta=1}^{1,M=5}&=\frac{1}{2\sqrt{5}}\left(|\psi\rangle_{T^2}^{0,5} +\frac{-1+3\sqrt{5}}{4} |\psi\rangle_{T^2}^{1,5}-\frac{1+\sqrt{5}}{4}|\psi\rangle_{T^2}^{2,5}-\frac{1+\sqrt{5}}{4}|\psi\rangle_{T^2}^{3,5}+\frac{-1+3\sqrt{5}}{4}|\psi\rangle_{T^2}^{4,5}\right)\,,
    \nonumber\\
    |\Psi\rangle_{\mathbb{Z}_4, \eta=1}^{2,M=5}&=\frac{1}{2\sqrt{5}}\left(|\psi\rangle_{T^2}^{0,5} -\frac{1+\sqrt{5}}{4} |\psi\rangle_{T^2}^{1,5}+\frac{-1+3\sqrt{5}}{4}|\psi\rangle_{T^2}^{2,5}+\frac{-1+3\sqrt{5}}{4}|\psi\rangle_{T^2}^{3,5}-\frac{1+\sqrt{5}}{4}|\psi\rangle_{T^2}^{4,5}\right)\,,
    \nonumber\\
    |\Psi\rangle_{\mathbb{Z}_4, \eta=\omega}^{M=5}&= -\frac{1}{40} \left(10 + \sqrt{10 \left(5 + \sqrt{5}\right)}\right)\left(|\psi\rangle_{T^2}^{1,5} - |\psi\rangle_{T^2}^{4,5}\right) -\frac{1}{8} \sqrt{2 - \frac{2}{\sqrt{5}}}\left(|\psi\rangle_{T^2}^{2,5} -|\psi\rangle_{T^2}^{3,5} \right)\,,  
    \nonumber\\
    |\Psi\rangle_{\mathbb{Z}_4, \eta=\omega^2}^{M=5}&=\frac{1}{2\sqrt{5}}\left(-|\psi\rangle_{T^2}^{0,5} +\frac{1+\sqrt{5}}{4} |\psi\rangle_{T^2}^{1,5}+\frac{1+\sqrt{5}}{4}|\psi\rangle_{T^2}^{2,5}+\frac{1+\sqrt{5}}{4}|\psi\rangle_{T^2}^{3,5}+\frac{1+\sqrt{5}}{4}|\psi\rangle_{T^2}^{4,5}\right)\,,
    \nonumber\\
    |\Psi\rangle_{\mathbb{Z}_4, \eta=\omega^3}^{M=5}&= -\frac{1}{40} \left(10 - \sqrt{10 \left(5 + \sqrt{5}\right)}\right)\left(|\psi\rangle_{T^2}^{1,5} - |\psi\rangle_{T^2}^{4,5}\right) +\frac{1}{8} \sqrt{2 - \frac{2}{\sqrt{5}}}\left(|\psi\rangle_{T^2}^{2,5} -|\psi\rangle_{T^2}^{3,5} \right)\,.  
\end{align}
From Eq. \eqref{eq:trfZ4}, we find the following transformations of chiral matters:
\small
\begin{align}
    \begin{pmatrix}
\frac{1}{4} \left(1 + \sqrt{5}\right) & \frac{1}{2} \left(-1 + \sqrt{5}\right) & 0 & \frac{1}{4} \left(15 - \sqrt{5}\right) & 0 \\
\frac{1}{4} \left(-3 + \sqrt{5}\right) & \frac{1}{2} \left(1 - \sqrt{5}\right) & 0 & \frac{1}{4} \left(-5 - \sqrt{5}\right) & 0 \\
0 & 0 & \frac{1}{4} \left(-1 + \sqrt{5} - 2 \sqrt{5 - 2 \sqrt{5}}\right) & 0 & c_1 \\
\frac{1}{4} \left(-5 - \sqrt{5}\right) & 0 & 0 & \frac{1}{4} \left(-1 - \sqrt{5}\right) & 0 \\
0 & 0 & c_2 & 0 & \frac{1}{4} \left(-1 + \sqrt{5} + 2 \sqrt{5 - 2 \sqrt{5}}\right)
    \end{pmatrix}
    \,,
\end{align}
\normalsize
for $n_1=n_2=1$ with
\small
\begin{align}
    c_1 &= \frac{5 \left(-8 + 4 \sqrt{5} + \sqrt{50 - 10 \sqrt{5}} - 3 \sqrt{10 - 2 \sqrt{5}} + 3 \sqrt{2 \left(5 + \sqrt{5}\right)} - \sqrt{10 \left(5 + \sqrt{5}\right)}\right)}{-100 + 44 \sqrt{5}}\,,
    \nonumber\\
    c_2&=\frac{1}{8} \left(-\sqrt{50 - 10 \sqrt{5}} - 3 \sqrt{10 - 2 \sqrt{5}} + 2 \left(-5 - 3 \sqrt{5} + 3 \sqrt{2 \left(5 + \sqrt{5}\right)} + \sqrt{10 \left(5 + \sqrt{5}\right)}\right)\right)\,,
\end{align}
\normalsize
and 
\small
\begin{align}
    \begin{pmatrix}
\frac{1}{2} \left(-1 + \sqrt{5}\right) & \frac{1}{4} \left(3 - \sqrt{5}\right) & 0 & \frac{1}{4} \left(5 + \sqrt{5}\right) & 0 \\
\frac{1}{2} \left(1 - \sqrt{5}\right) & \frac{1}{4} \left(-1 - \sqrt{5}\right) & 0 & \frac{1}{4} \left(-15 + \sqrt{5}\right) & 0 \\
0 & 0 & \frac{1}{4} \left(1 - \sqrt{5} - 2 \sqrt{5 - 2 \sqrt{5}}\right) & 0 & \frac{1}{4} \left(5 + 3 \sqrt{5} + 2 \sqrt{5 \left(5 + 2 \sqrt{5}\right)}\right) \\
0 & \frac{1}{4} \left(5 + \sqrt{5}\right) & 0 & \frac{1}{4} \left(1 + \sqrt{5}\right) & 0 \\
0 & 0 & c & 0 & \frac{1}{4} \left(1 - \sqrt{5} + 2 \sqrt{5 - 2 \sqrt{5}}\right)
    \end{pmatrix}
    \,,
\end{align}
\normalsize
for $n_1=n_2=2$ with 
\small
\begin{align}
    c_3= \frac{1}{8} \left(10 + 6 \sqrt{5} + \sqrt{50 - 10 \sqrt{5}} + 3 \sqrt{10 - 2 \sqrt{5}} - 6 \sqrt{2 \left(5 + \sqrt{5}\right)} - 2 \sqrt{10 \left(5 + \sqrt{5}\right)}\right)\,.
\end{align} 
\normalsize
Here, we numerically show the value of representation matrix. 
The representation matrices in $n_1=n_2=3$ and $n_1=n_2=4$ cases correspond to the sign flipping of those in $n_1=n_2=2$ and $n_1=n_2=1$, respectively. 
In any case, one can realize the invertible transformations obeying the fusion rule \eqref{eq:fusionrulesZN}.

\paragraph{$M=6$}\,\\

When $M=6$, the numbers of $\mathbb{Z}_4$-untwisted and -twisted states are shown in Table~\ref{tab:Z4}. 
By diagonalizing $ (M_\eta)_k^j$, 
zero-mode wave functions for each $\mathbb{Z}_4$ eigenvalue are given by
\begin{align}
    |\Psi\rangle_{\mathbb{Z}_4, \eta=1}^{1,M=6}
    &=\frac{1}{2\sqrt{6}}\left(|\psi\rangle_{T^2}^{0,6} +\frac{1+\sqrt{6}}{2} |\psi\rangle_{T^2}^{1,6}-\frac{1}{2}|\psi\rangle_{T^2}^{2,6}-|\psi\rangle_{T^2}^{3,6}-\frac{1}{2}|\psi\rangle_{T^2}^{4,6}+\frac{1+\sqrt{6}}{2} |\psi\rangle_{T^2}^{5,6}\right)\,,
    \nonumber\\
    |\Psi\rangle_{\mathbb{Z}_4, \eta=1}^{2,M=6}&=\frac{1}{2\sqrt{6}}\left(|\psi\rangle_{T^2}^{0,6} -\frac{1}{2}|\psi\rangle_{T^2}^{1,6}-\frac{1-\sqrt{6}}{2} |\psi\rangle_{T^2}^{2,6}+
    |\psi\rangle_{T^2}^{3,6}-\frac{1-\sqrt{6}}{2} |\psi\rangle_{T^2}^{4,6}-\frac{1}{2} |\psi\rangle_{T^2}^{5,6}\right)\,,
    \nonumber\\
    |\Psi\rangle_{\mathbb{Z}_4, \eta=\omega^2}^{M=6}
    &=\frac{1}{4\sqrt{2}}\left(-(1+\sqrt{2})|\psi\rangle_{T^2}^{1,6} -|\psi\rangle_{T^2}^{2,5}+|\psi\rangle_{T^2}^{4,6}+(1+\sqrt{2})|\psi\rangle_{T^2}^{5,6}\right)\,,
        \nonumber\\
    |\Psi\rangle_{\mathbb{Z}_4, \eta=\omega^2}^{1,M=6}
    &=\frac{1}{2\sqrt{6}}\left(-|\psi\rangle_{T^2}^{0,6} +\frac{-1+\sqrt{6}}{2} |\psi\rangle_{T^2}^{1,6}+\frac{1}{2}|\psi\rangle_{T^2}^{2,6}+|\psi\rangle_{T^2}^{3,6}+\frac{1}{2}|\psi\rangle_{T^2}^{4,6}+\frac{-1+\sqrt{6}}{2} |\psi\rangle_{T^2}^{5,6}\right)\,,
    \nonumber\\
    |\Psi\rangle_{\mathbb{Z}_4, \eta=\omega^2}^{2,M=6}&=\frac{1}{2\sqrt{6}}\left(-|\psi\rangle_{T^2}^{0,6} +\frac{1}{2}|\psi\rangle_{T^2}^{1,6} +\frac{1+\sqrt{6}}{2} |\psi\rangle_{T^2}^{2,6}-
    |\psi\rangle_{T^2}^{3,6}+\frac{1+\sqrt{6}}{2} |\psi\rangle_{T^2}^{4,6}+\frac{1}{2} |\psi\rangle_{T^2}^{5,6}\right)\,,
    \nonumber\\
    |\Psi\rangle_{\mathbb{Z}_4, \eta=\omega^3}^{M=6}
    &=\frac{1}{4\sqrt{2}}\left((1-\sqrt{2})|\psi\rangle_{T^2}^{1,6} +|\psi\rangle_{T^2}^{2,6} -|\psi\rangle_{T^2}^{4,6}+(-1+\sqrt{2})|\psi\rangle_{T^2}^{5,6}\right)\,,
\end{align}
From Eq. \eqref{eq:trfZ4}, we find the following transformations
\small
\begin{align}
    \begin{pmatrix}
   |\Psi\rangle_{\mathbb{Z}_4, \eta=1}^{1,M=6}\\ 
   |\Psi\rangle_{\mathbb{Z}_4, \eta=1}^{2,M=6}\\ 
   |\Psi\rangle_{\mathbb{Z}_4, \eta=\omega}^{M=6}\\ 
   |\Psi\rangle_{\mathbb{Z}_4, \eta=\omega^2}^{1,M=6}\\
   |\Psi\rangle_{\mathbb{Z}_4, \eta=\omega^2}^{2,M=6}\\
   |\Psi\rangle_{\mathbb{Z}_4, \eta=\omega^3}^{M=6}\\
    \end{pmatrix}
    \rightarrow 
    \frac{1}{2\sqrt{2}}
    \begin{pmatrix}
3 & 3 + \sqrt{6} & 0 & 3 & 3 + \sqrt{6} & 0 \\
3 - \sqrt{6} & -3 & 0 & 3 - \sqrt{6} & -3 & 0 \\
0 & 0 & 0 & 0 & 0 & -2 \sqrt{3} (2 + \sqrt{2}) \\
-3 & -3 + \sqrt{6} & 0 & -3 & -3 + \sqrt{6} & 0 \\
-3 - \sqrt{6} & 3 & 0 & -3 - \sqrt{6} & 3 & 0 \\
0 & 0 & \frac{4 \sqrt{3}}{2 + \sqrt{2}} & 0 & 0 & 0 \\
    \end{pmatrix}
    \begin{pmatrix}
   |\Psi\rangle_{\mathbb{Z}_4, \eta=1}^{1,M=6}\\ 
   |\Psi\rangle_{\mathbb{Z}_4, \eta=1}^{2,M=6}\\ 
   |\Psi\rangle_{\mathbb{Z}_4, \eta=\omega}^{M=6}\\ 
   |\Psi\rangle_{\mathbb{Z}_4, \eta=\omega^2}^{1,M=6}\\
   |\Psi\rangle_{\mathbb{Z}_4, \eta=\omega^2}^{2,M=6}\\
   |\Psi\rangle_{\mathbb{Z}_4, \eta=\omega^3}^{M=6}\\
    \end{pmatrix}
    \,,
\end{align}
\normalsize
for $n_1=n_2=1,5$, and
\small
\begin{align}
    \begin{pmatrix}
   |\Psi\rangle_{\mathbb{Z}_4, \eta=1}^{1,M=6}\\ 
   |\Psi\rangle_{\mathbb{Z}_4, \eta=1}^{2,M=6}\\ 
   |\Psi\rangle_{\mathbb{Z}_4, \eta=\omega}^{M=6}\\ 
   |\Psi\rangle_{\mathbb{Z}_4, \eta=\omega^2}^{1,M=6}\\
   |\Psi\rangle_{\mathbb{Z}_4, \eta=\omega^2}^{2,M=6}\\
   |\Psi\rangle_{\mathbb{Z}_4, \eta=\omega^3}^{M=6}\\
    \end{pmatrix}
    \rightarrow 
    \frac{\sqrt{3}}{2\sqrt{2}}
    \begin{pmatrix}
-1 - \sqrt{\frac{2}{3}} & 1 & 0 & 3 + \sqrt{6} & -3 & 0 \\
1 & 1 - \sqrt{\frac{2}{3}} & 0 & -3 & -3 + \sqrt{6} & 0 \\
0 & 0 & 2 \sqrt{\frac{2}{3}} & 0 & 0 & 0 \\
-3 + \sqrt{6} & 3 & 0 & 1 - \sqrt{\frac{2}{3}} & -1 & 0 \\
3 & 3 + \sqrt{6} & 0 & -1 & -1 - \sqrt{\frac{2}{3}} & 0 \\
0 & 0 & 0 & 0 & 0 & 2 \sqrt{\frac{2}{3}} \\
    \end{pmatrix}
    \begin{pmatrix}
   |\Psi\rangle_{\mathbb{Z}_4, \eta=1}^{1,M=6}\\ 
   |\Psi\rangle_{\mathbb{Z}_4, \eta=1}^{2,M=6}\\ 
   |\Psi\rangle_{\mathbb{Z}_4, \eta=\omega}^{M=6}\\ 
   |\Psi\rangle_{\mathbb{Z}_4, \eta=\omega^2}^{1,M=6}\\
   |\Psi\rangle_{\mathbb{Z}_4, \eta=\omega^2}^{2,M=6}\\
   |\Psi\rangle_{\mathbb{Z}_4, \eta=\omega^3}^{M=6}\\
    \end{pmatrix}
    \,.
\end{align}
\normalsize
for $n_1=n_2=2,4$. For $n_1=n_2=3$, we find that the transformation is described by a trivial representation $\hat{U}_{\mathbb{Z}_4}^{(\lambda_1, \lambda_2)}=0$ with $n_1=n_2=3$. 
In any case, the representation matrices of chiral zero modes obey the fusion rule \eqref{eq:fusionrulesZN}.

\subsection{$T^2/\mathbb{Z}_6$}
\label{app:Z6}

For $M=2$, the chiral zero modes and their transformations under the non-invertible symmetry are presented in Sec. \ref{sec:Z6}. 
Hence, we show $M=4,6$ cases in the following. 

\paragraph{$M=4$}\,\\

When $M=4$, the numbers of $\mathbb{Z}_6$-untwisted and -twisted states are shown in Table~\ref{tab:Z6}. 
By diagonalizing $ (M_\eta)_k^j$, 
zero-mode wave functions for each $\mathbb{Z}_6$ eigenvalue are given by
\small
\begin{align}
 |\Psi\rangle_{\mathbb{Z}_6, \eta=1}^{M=4} 
 &=\frac{1}{6}\left( \left(e^{\frac{\pi i}{12}} + e^{\frac{\pi i}{6}}\right)|\psi\rangle_{T^2}^{0,4} 
 + |\psi\rangle_{T^2}^{1,4} + \left(e^{\frac{\pi i}{12}} - e^{\frac{\pi i}{6}}\right)|\psi\rangle_{T^2}^{2,4}
 + |\psi\rangle_{T^2}^{3,4}
 \right)    \,,
    \nonumber\\
|\Psi\rangle_{\mathbb{Z}_6, \eta=\omega}^{M=4} &= 
 \frac{1}{2} \left( -|\psi\rangle_{T^2}^{1,4} +|\psi\rangle_{T^2}^{3,4}\right) 
    \,,
    \nonumber\\
|\Psi\rangle_{\mathbb{Z}_6, \eta=\omega^2}^{M=4} 
 &=\frac{1}{6}\left( \frac{(4+(1+3i)\sqrt{2}-4\sqrt{3}i - (1+i)\sqrt{6}}{2(i-\sqrt{3}+2e^{\frac{\pi i}{4}})}|\psi\rangle_{T^2}^{0,4} 
 + |\psi\rangle_{T^2}^{1,4} + \frac{2 + \sqrt{3} +i}{i-\sqrt{3}+2e^{\frac{3\pi i}{4}}}|\psi\rangle_{T^2}^{2,4}
 + |\psi\rangle_{T^2}^{3,4}
 \right)    \,,
    \nonumber\\
|\Psi\rangle_{\mathbb{Z}_3, \eta=\omega^2}^{M=4} 
 &=\frac{1}{6}\left( \left(-i  + e^{\frac{3\pi i}{4}}\right)|\psi\rangle_{T^2}^{0,4} 
 + |\psi\rangle_{T^2}^{1,4} + \left(i + e^{\frac{3\pi i}{4}}\right)|\psi\rangle_{T^2}^{2,4}
 + |\psi\rangle_{T^2}^{3,4}
 \right)    \,.
 \end{align}
\normalsize

We find that under the non-invertible symmetry, the transformation matrix of chiral zero modes is given by
\small
\begin{align}
    \begin{pmatrix}
-\frac{2}{3} \sqrt{(-3 - 4 i) (-2 + \sqrt{3})} & 0 & \frac{4}{3} (-1)^{5/12} & \frac{2}{3} \sqrt{(3 + 2 i) + (2 + i) \sqrt{3}} \\
0 & 0 & 0 & 0 \\
-\frac{4}{3} (-1)^{1/12} & 0 & \left(\frac{4}{3} + \frac{2 i}{3}\right) \sqrt{2 + \sqrt{3}} & \frac{2}{3} \sqrt{(3 + 2 i) - (2 + i) \sqrt{3}} \\
-\frac{2}{3} \sqrt{(-3 + 2 i) - (2 - i) \sqrt{3}} & 0 & -\frac{2}{3} \sqrt{(-3 + 2 i) + (2 - i) \sqrt{3}} & \left(-\frac{4}{3} - \frac{2 i}{3}\right) \sqrt{2} \\
    \end{pmatrix}
    \,,
\end{align}
\normalsize
for $n_1=n_2=1,3$, and
\small
\begin{align}
 2
    \begin{pmatrix}
1 & 0 & 0 & 0 \\
0 & -3 & 0 & 0 \\
0 & 0 & 1 & 0 \\
0 & 0 & 0 & 1 \\
    \end{pmatrix}
    \,.
\end{align}
\normalsize
for $n_1=n_2=2$. 
Here, we present show the transformation matrix in the basis of 
\begin{align}
\{|\Psi\rangle_{\mathbb{Z}_6, \eta=1}^{M=4}, |\Psi\rangle_{\mathbb{Z}_6, \eta=\omega}^{M=4},|\Psi\rangle_{\mathbb{Z}_6, \eta=\omega^2}^{M=4},|\Psi\rangle_{\mathbb{Z}_6, \eta=\omega^4}^{M=4}\}\,.
\end{align}
One can obtain the similar transformations for other $\{n_1,n_2\}$, 
and the fusion rule \eqref{eq:fusionrulesZN} is satisfied. 

\paragraph{$M=6$}\,\\

When $M=6$, the numbers of $\mathbb{Z}_3$-untwisted and -twisted states are shown in Table~\ref{tab:Z6}. 
By diagonalizing $ (M_\eta)_k^j$, 
zero-mode wave functions for each $\mathbb{Z}_6$ eigenvalue are given by
\begin{align}
 |\Psi\rangle_{\mathbb{Z}_6, \eta=1}^{1,M=6} 
 &\simeq (-0.0327002 - 0.34663 i) |\psi\rangle_{T^2}^{0,6} 
 +0.285699|\psi\rangle_{T^2}^{1,6} -(0.41285 - 0.074982 i)|\psi\rangle_{T^2}^{2,6}
\nonumber\\
&+(0.0101232 + 0.602738 i)|\psi\rangle_{T^2}^{3,6}
-(0.41285 - 0.074982 i)|\psi\rangle_{T^2}^{4,6}
+0.285699|\psi\rangle_{T^2}^{5,6}
    \,,
    \nonumber\\
|\Psi\rangle_{\mathbb{Z}_6, \eta=1}^{2,M=6} 
&\simeq 
(0.585094 + 0.364238 i) |\psi\rangle_{T^2}^{0,6} 
 +0.465283|\psi\rangle_{T^2}^{1,6} +(0.150099 + 0.0573633 i)|\psi\rangle_{T^2}^{2,6}
\nonumber\\
&-(0.200594 - 0.0118966 i)|\psi\rangle_{T^2}^{3,6}
+(0.150099 + 0.0573633 i)|\psi\rangle_{T^2}^{4,6}+0.465283|\psi\rangle_{T^2}^{5,6}
    \,,
    \nonumber\\
|\Psi\rangle_{\mathbb{Z}_6, \eta=\omega}^{M=6} 
&\simeq 
-0.627963 |\psi\rangle_{T^2}^{1,6} 
-(0.22985 - 0.22985 i)|\psi\rangle_{T^2}^{2,6} 
 \nonumber\\
&+(0.22985 - 0.22985 i) |\psi\rangle_{T^2}^{4,6}
+0.627963|\psi\rangle_{T^2}^{5,6}
   \,,
    \nonumber\\
|\Psi\rangle_{\mathbb{Z}_6, \eta=\omega^2}^{M=6} 
&\simeq 
0.51273 |\psi\rangle_{T^2}^{0,6} 
-(0.256365 - 0.0686928 i)|\psi\rangle_{T^2}^{1,6} -(0.256365 - 0.444037 i)|\psi\rangle_{T^2}^{2,6}
 \nonumber\\
&+(0.187672 + 0.187672 i)|\psi\rangle_{T^2}^{3,6}
-(0.256365 - 0.444037 i)|\psi\rangle_{T^2}^{4,6}
\nonumber\\
&-(0.256365 - 0.0686928 i) |\psi\rangle_{T^2}^{5,6}
   \,,
    \nonumber\\
|\Psi\rangle_{\mathbb{Z}_6, \eta=\omega^3}^{M=6} 
&\simeq 
-0.325058 |\psi\rangle_{T^2}^{1,6} 
+(0.444037 - 0.444037 i)|\psi\rangle_{T^2}^{2,6}
 \nonumber\\
&-(0.444037 - 0.444037 i) |\psi\rangle_{T^2}^{4,6}
+0.325058 |\psi\rangle_{T^2}^{5,6}
   \,,
    \nonumber\\
|\Psi\rangle_{\mathbb{Z}_6, \eta=\omega^4}^{M=6} 
&\simeq 
(-0.265408 + 0.265408 i) |\psi\rangle_{T^2}^{0,6} 
+(0.181277 - 0.313982 i)|\psi\rangle_{T^2}^{1,6} 
\nonumber\\
&+(0.0485731 + 0.181277 i)|\psi\rangle_{T^2}^{2,6}
+0.725109|\psi\rangle_{T^2}^{3,6}
+(0.0485731 + 0.181277 i)|\psi\rangle_{T^2}^{4,6}
\nonumber\\
&+(0.181277 - 0.313982 i) |\psi\rangle_{T^2}^{5,6}
\,.
\end{align}

We find that under the non-invertible symmetry, the transformation matrix of chiral zero modes is given by
\small
 \begin{align}
 \begin{pmatrix}
-0.446 + 0.773i & -0.235 + 0.165i & 0 & -0.004 - 2.414i & 0 & 0.385 - 1.031i \\
-0.026 + 0.286i & -0.420 + 0.727i & 0 & -0.929 - 1.994i & 0 & 0.093 + 1.203i \\
0 & 0 & 0.289 - 0.500i & 0 & -0.816 - 1.414i & 0 \\
2.092 - 1.204i & 2.192 - 0.192i & 0 & 0.577 - 1.000i & 0 & 0 \\
0 & 0 & 1.633 & 0 & -0.289 + 0.500i & 0 \\
0.700 - 0.849i & -1.089 + 0.521i & 0 & 0 & 0 & 0.289 - 0.500i
    \end{pmatrix}
\,,
 \end{align}
\normalsize
for $n_1=n_2=1,5$, 
\small
 \begin{align}
 \begin{pmatrix}
-0.958 + 1.659i & -3.666 + 2.576i & 0 & 0 & 0 & 0 \\
-0.398 + 4.463i & -0.542 + 0.939i & 0 & 0 & 0 & 0 \\
0 & 0 & 1.500 - 2.598i & 0 & 0 & 0 \\
0 & 0 & 0 & -3.000 + 5.196i & 0 & 0 \\
0 & 0 & 0 & 0 & 1.500 - 2.598i & 0 \\
0 & 0 & 0 & 0 & 0 & 1.500 - 2.598i
    \end{pmatrix}
\,,
 \end{align}
\normalsize
for $n_1=n_2=2,4$, and 
\small
 \begin{align}
 \begin{pmatrix}
0.107 & 1.043 + 0.484i & 0 & -2.088 + 1.210i & 0 & 2.170 - 0.364i \\
1.043 - 0.484i & -0.107 & 0 & -1.263 + 1.802i & 0 & -1.991 + 1.364i \\
0 & 0 & 1.155 & 0 & 1.633 - 2.828i & 0 \\
-2.088 - 1.210i & -1.263 - 1.802i & 0 & -1.155 & 0 & 0 \\
0 & 0 & 1.633 + 2.828i & 0 & -1.155 & 0 \\
2.170 + 0.364i & -1.991 - 1.364i & 0 & 0 & 0 & 1.155
    \end{pmatrix}
 \,,
 \end{align}
\normalsize
for $n_1=n_2=3$.

Here, we numerically show the transformation matrix in the basis of 
\begin{align}
\{|\Psi\rangle_{\mathbb{Z}_6, \eta=1}^{1,M=6}, |\Psi\rangle_{\mathbb{Z}_6, \eta=1}^{2,M=6},|\Psi\rangle_{\mathbb{Z}_6, \eta=1}^{3,M=6},|\Psi\rangle_{\mathbb{Z}_3, \eta=\omega}^{1,M=6},|\Psi\rangle_{\mathbb{Z}_3, \eta=\omega}^{M=6},|\Psi\rangle_{\mathbb{Z}_6, \eta=\omega^2}^{M=6} ,|\Psi\rangle_{\mathbb{Z}_6, \eta=\omega^3}^{M=6} ,|\Psi\rangle_{\mathbb{Z}_6, \eta=\omega^4}^{M=6}\}.
\end{align}
One can obtain the similar transformations for other $\{n_1,n_2\}$, and the fusion rule \eqref{eq:fusionrulesZN} is satisfied.

\bibliography{references}{}
\bibliographystyle{JHEP}

\end{document}